\documentclass[onecolumn, numberedappendix]{aastex6}
\usepackage{graphicx}
\usepackage{epsfig}
\usepackage{natbib}
\usepackage{amssymb}
\usepackage{amsbsy}
\usepackage{natbib}
\usepackage{url}
\usepackage[mathcal]{euscript}
\bibpunct{(}{)}{,}{a}{}{,}

\newcommand{\etal}{et~al.}

\begin{document}
 
\def\simlt{\vcenter{\hbox{$<$}\offinterlineskip\hbox{$\sim$}}}
\def\simgt{\vcenter{\hbox{$>$}\offinterlineskip\hbox{$\sim$}}}
\def\etal{et al.\ }
\def\kms{km s$^{-1}$}

\title{More Rapidly Rotating PMS M Dwarfs with Light Curves Suggestive of Orbiting Clouds of Material}
\author{John Stauffer\altaffilmark{1},
Luisa Rebull\altaffilmark{1,3},
Trevor J. David\altaffilmark{2},
Moira Jardine\altaffilmark{5},
Andrew Collier Cameron\altaffilmark{5},
Ann Marie Cody\altaffilmark{4},
Lynne A. Hillenbrand\altaffilmark{2},
David Barrado\altaffilmark{6},
Julian van Eyken\altaffilmark{7},
Carl Melis\altaffilmark{8},
Cesar Briceno\altaffilmark{9}.
}
\altaffiltext{1}{Spitzer Science Center (SSC), IPAC, California Institute of
Technology, Pasadena, CA 91125, USA}
\altaffiltext{2}{Astronomy Department,
California Institute of Technology, Pasadena, CA 91125 USA}
\altaffiltext{3}{Infrared Science Archive (IRSA), IPAC, 1200 E. California Blvd,
MS 314-6, California Institute of Technology, Pasadena, CA 91125 USA}
\altaffiltext{4}{NASA Ames Research Center, Space Sciences and
Astrobiology Division, MS245-3, Moffett Field, CA 94035 USA}
\altaffiltext{5}{School of Physics and Astronomy, University of St. Andrews, North
Haugh, St Andrews KY16922, UK}
\altaffiltext{6}{Centro de Astrobiolog\'ia, Dpto. de
Astrof\'isica, INTA-CSIC, E-28692, ESAC Campus, Villanueva de
la Ca\~nada, Madrid, Spain}
\altaffiltext{7}{NASA Exoplanet Science Institute,
California Institute of Technology, Pasadena, CA 91125 USA}
\altaffiltext{8}{Center for Astrophysics and Space Sciences, University of California, San Diego, CA 92093-0424, USA}
\altaffiltext{9}{Cerro Tololo Inter-American Observatory, Casilla 603, La Serena 1700000, Chile 0000-0001-7124-4094}

\email{stauffer@ipac.caltech.edu}

\begin{abstract}

In a previous paper, using data from K2 Campaign 2, we 
identified 11 very low mass members of the
$\rho$ Oph and Upper Scorpius star-forming region as having periodic
photometric variability and phased light curves showing multiple
scallops or undulations.  All the stars with the ``scallop-shell"
light curve morphology are mid-to-late M
dwarfs without evidence of active accretion, and with photometric
periods generally $<$1 day.  Their phased light curves have too much
structure to be attributed to non-axisymmetrically distributed photospheric
spots and rotational modulation.  We have now identified an additional
eight probable members of the same star-forming region plus three 
stars in the Taurus star-forming region with this same light
curve morphology
and sharing the same period and spectral type range as the
previous group.   We describe the light curves of these new stars in
detail and present their general physical characteristics.   We also
examine the properties of the overall set of stars in order to
identify common features that might help elucidate the causes of their
photometric variability.

\end{abstract}

\section{Introduction}

If observed with enough precision, nearly all
stars show some type of short term photometric variability,
with many different physical mechanisms as the source of that
variability.  NASA's Kepler mission, and its successor K2 mission (Howell \etal\ 2014),
have produced long duration, continuous, high precision light curves
of more than 300,000 stars which have served to highlight many of
these known mechanisms for producing photometric variability.
Inevitably, the quality and quantity of those light curves has also
resulted in identification of stars whose photometric variability is
not easily associated with any of the previously known physical
mechanisms and which offer the possibility for exploring new
physics.   In Stauffer et al. (2017; hereafter S17), we identified 23
stars observed by K2  in the $\sim$8 Myr old Upper Scorpius
association as having photometric variability not easily attributable
to the mechanisms usually associated with low mass, pre-main sequence (PMS) 
stars. All of these stars are mid to late M dwarfs with no obvious IR
excess nor other indication of ongoing accretion; most are very rapid
rotators with periods often less than 1 day.   Half of those stars
fall in a category which we dubbed ``scallop-shell" light curves.  
These latter stars all had P $<$ 0.65 days and very structured phased light
curves with many humps or arcs and variability amplitudes sometimes up
to 10\%.   For about half these stars, the light curve shape is
remarkably stable over the 78 day duration of the campaign; for the
other half, there are one or two sudden state-changes in the
morphology of the phased light curve usually only affecting about 20\%
of the waveform and often occurring coincident in time with a major
flare event.  S17 and David \etal\ (2017) also describe two other types
of rapidly rotating M dwarfs with differing unusual light curve
morphologies.  We do not address these latter types of variability 
in this paper, though it
may be that in some cases the physical mechanism driving the variability
may be the same as for the stars with scallop-shell light curve morphology.

We have now identified eight more low mass members of the Upper
Sco association or its younger neighbor, the $\rho$ Oph cluster (age
$\sim$ 1 Myr), whose K2 Campaign 2 light curves  are unusual and in many cases
share the same characteristics as the scallop-shell stars in S17.  From
K2 Campaign 13, we
have also identified three low mass members of the Taurus star-forming
region that share those characteristics as well.   In \S
2, we describe the young stars targeted in Campaigns 2 and 13 of the K2
mission and the  sources of data for this paper; we also provide a
table identifying the stars with unusual light curves which are the
topic of this paper.   In \S 3, we illustrate the light curve
morphologies for the new set of stars with ``scallop-shell" light
curves.  In \S 4, we discuss the light curve of one particular star,
EPIC 204099739, whose pedigree is less certain.  in \S 5, we review
the properties of all the stars that have light curves with this
morphology.

\section{Newly Identified Stars of Interest and Observational Data}

The stars we discuss in this paper were observed by NASA's K2 mission
during its Campaign 2 (August to November 2014) or its Campaign 13
(March to May 2017).  The light curves we examined were all associated
with stars identified as probable or possible members of the Upper
Sco and $\rho$ Oph star-forming region (Campaign 2) or the Taurus
star-forming region (Campaign 13).  The
original set of stars with unusual light curve morphologies discussed
in S17 were identified during the early stages of our project to measure
rotation periods and accretion signatures for the PMS stars in Upper Sco
when both our membership lists and our light curve processing were
at preliminary stages.  We now have an improved
membership list, more refined light curve versions, and a more
complete set of rotational data (Rebull \etal\ 2018).   In the process
of reviewing the final set of light curves, we identified 8
additional rapidly rotating M dwarfs with unusual,  structured, periodic
light curves similar in morphology to the stars with ``scallop shell"
light curve morphologies in S17.   In addition,
from the Campaign 13 data, we have identified three low mass candidate
Taurus members with the same properties.
All of these stars are listed in Table 1, where we also provide some
basic photometric and spectroscopic properties for these stars.

\floattable
\begin{deluxetable*}{lccccccccccccc}
\tabletypesize{\footnotesize}
\tablecolumns{12}
\tablewidth{0pt}
\tablecaption{Low Mass Pre-Main Sequence Stars with Unusual Periodic Light Curves\label{tab:basicdata}}
\tablehead{
\colhead{EPIC ID } &
\colhead{RA } &
\colhead{Dec} &
\colhead{$K_s$} &
\colhead{$V-K_s$ \tablenotemark{a}} &
\colhead{($V-K_s$)$_o$}  &
\colhead{$[W1]-$[W3]} &
\colhead{SpT\tablenotemark{b}} &
\colhead{H$\alpha$ EqW} &
\colhead{Li EqW} &
\colhead{P1\tablenotemark{c}} &
\colhead{P2\tablenotemark{c}}  \\
 & \colhead{(deg)}& \colhead{(deg)}
& \colhead{(mag)}& \colhead{(mag)}
& \colhead{(mag)} & \colhead{(mag)}
& &\colhead{(\AA)} &\colhead{(\AA)}
& \colhead{(days)}& \colhead{(days)}  }
\startdata
\cutinhead{Upper Sco Scallop shells}
203354381  & 246.616 & -26.421 & 9.845 &  5.88 & 5.24 &  0.200 & \nodata & \nodata  & \nodata  & 0.5993 & \nodata \\
203636498 & 245.271 & -25.462 & 10.943 & 6.34  & 6.06 & 0.310: & M5.5   &  -9.3 & 0. 61 & 0.7794 & \nodata \\
204060981 & 242.734 & -23.825 &  9.473 &  5.88 & 5.11 &  0.279 &  M4.5   & -6.7  &  0.65 & 0.3996* & 0.3802* \\
205267399 & 244.674 & -18.544 & 10.082 & 5.52 & 5.17  & 0.272 &   M5    &  -9.0 & no:  & 0.3311 & 0.3344* \\
\cutinhead{$\rho$ Oph Scallop shells}
203821589  & 247.000 & -24.805 & 9.269 & 8.15 & 5.66 & 0.470 &  M4.75  & -7.5   & yes: & 0.9105 & 0.6677*  \\
203897692  & 246.484 & -24.504 & 9.764 & $>$10 & 6.03  & \nodata &   M5    &  -12  & \nodata      & 0.5011 & 0.6043*  \\
204185983  & 246.470 & -23.326 &  7.870 & 6.75 & 4.06  & 0.244 &  M1.5   & -2.5  & 0.68 & 1.0529 & \nodata \\
\cutinhead{Taurus Scallop shells}
246676629 &  74.868  & +14.273 &  10.062 & 5.46 & 5.26 & 0.292 & M3.5  & -5.9 & 0.4: & 0.6253* & 0.6332 \\
246682490 &  74.770  & +14.349 &   9.467 & 5.17 & 4.77 & 0.190 & M3.5 & -6.1 & 0.2: & 3.6324 & 0.4377* \\
247343526 &  70.223 &  +20.930 &  11.621 & 6.42 & 6.42 & 0.412 &  M5     & -11.0 & \nodata &  0.3568 & \nodata \\
\cutinhead{Possible Scallop shell in Upper Sco}
204099739 & 242.163 & -23.668 &  9.146 & 5.33 & 4.78  & 0.246 &  M3.5   & -5.0  & 0.56 & 0.7158 & 0.7428* \\
\enddata
\tablenotetext{a}{Observed $V-K_s$ color where both bands have published measurements in the literature.  
Otherwise, $V-K_s$ is estimated using a conversion based on the Gaia DR1 $G$ magnitudes
and 2MASS $K_s$ values.  }
\tablenotetext{b}{Spectral types and H$\alpha$\ equivalent widths for EPIC 203821589 and
203897692 are from Wilking \etal\ 2015, and for EPIC 247343526 are from Slesnick \etal\ 2006.
All other spectroscopic information are from our own spectra, as reported in the Appendix.}
\tablenotetext{c}{An asterisk is attached to the period (P1 or P2) to
  designate which star in a binary has the unusual light curve which
  is discussed in this paper.  All of the binary companions have
  normal, spotted-star light curve morphologies, except EPIC
  204060981, where both stars in the binary system have scallop-shell
  light curves.  EPIC 246682490 actually appears to be a triple; the third period is 0.4214d.}
\end{deluxetable*}
\noindent

The EPIC catalog available at MAST provides accurate coordinates for
all of the stars in our sample.   Using those coordinates, we
downloaded all available near and mid-IR photometry for our stars from
the 2MASS (Skrutskie \etal\ 2006), WISE (Wright \etal\ 2010), Spitzer
(Werner \etal\ 2004)
SEIP\footnote{http://irsa.ipac.caltech.edu/data/SPITZER/Enhanced/SEIP/overview.html}
and FEPS (Meyer \etal\ 2006),  AKARI (Murakami \etal\ 2007), and SDSS
(e.g., Ahn \etal\ 2014) archives as well as $B$ and $V$ band
photometry from APASS (Henden \& Munari 2014).   We also downloaded
$grizy$ photometry for most of our stars from the PAN-STARRS1 database
(Flewelling \etal\ 2016), and Gaia DR1 $G$ magnitudes from the IRSA
mirror site for Gaia DR1. Spectral types were available in the
literature for four of our stars  (Wilking \etal\ 2005, Slesnick \etal\ 2006,
Luhman \& Mamajek 2012); in many cases, those same
references also provided H$\alpha$ and lithium (6708\AA) equivalent
widths. We have used our own Keck HIRES, Palomar 5m DBSP, Lick 3m KAST, and
SOAR Goodman spectra to
provide spectral data for seven of the stars in Table 1, as described in the
Appendix.   We used all of the photometry and spectral type data to
produce spectral energy distributions (SEDs) for the stars in Table
1.    As was true for the original S17 sample, all of the new
stars appear to be mid-to-late M dwarfs, and
all appear not to have a detected IR excess.  The SEDs for
the new sample are provided in the Appendix.  

\begin{figure}[ht]
\epsscale{0.9}
\plotone{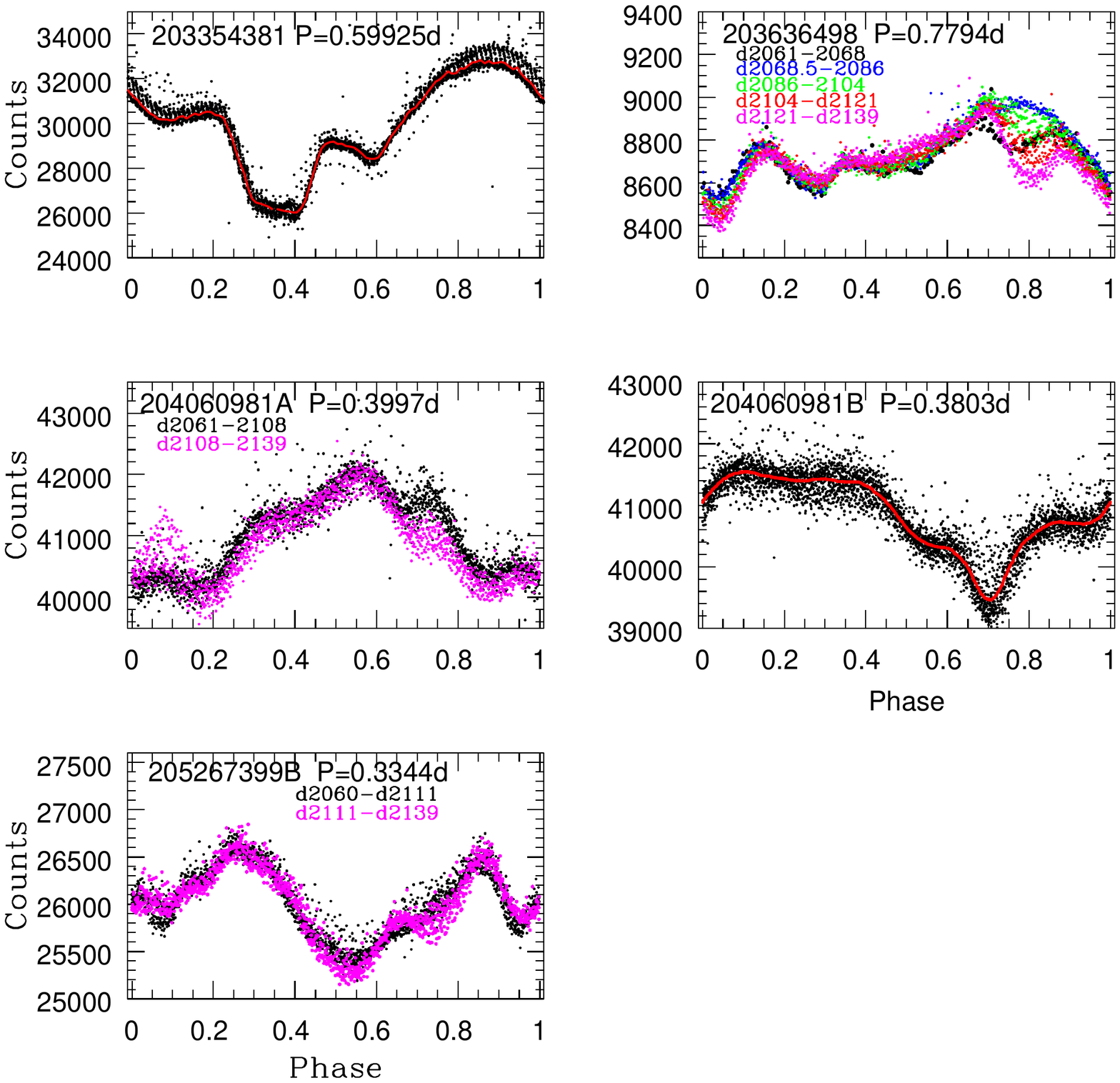}
\caption{The phased light curve  of the Upper Sco members in
Table 1 with scallop-shell phased light curves.
For EPIC 203636498, 204060981A and 205267e99B, the light curve shape evolved
considerably in a restricted phase interval over the course of the
campaign; for these stars, the light curve is broken up into two or more
time intervals, with each interval plotted as a different color, with
black being the first time interval and magenta the last.  
For the two
stars with no obvious changes in their morphology over the campaign,
we also show a median fit to the light
curve to ``guide the eye".
\label{fig:superfigure1}}
\end{figure}

At the time of the writing of this paper, light curves for the stars
in Campaign 2 were publicly available at MAST from Luger \etal\ (2017)
and Vanderburg \etal\ (2016).   We also had available to us two sets
of light curves created by co-author Ann Marie Cody (see Cody \etal\
2017, 2018).   For Campaign 13, we had from MAST the K2 project-provided
light curves (their PDC version) and the Vanderburg \etal\ (2016) light
curves, plus
our own light curves from co-author Ann Marie Cody.  
For each star, we examined all of the available light
curves, and chose the one which seemed to produce the final phased
light curve with the fewest artifacts and the highest
signal-to-noise.   In some cases, we did further processing on the
light curves prior to phasing to the period identified in the
Lomb-Scargle periodogram, either to remove long-term trends or to
separate the light of the two stars in a binary (that is, when the
periodogram showed two well-resolved peaks).   For these latter stars,
we list both periods in Table 1. It is likely that the two components
of these binaries have similar  masses, because otherwise the
secondary would be unlikely to contribute enough of the light to yield
a detectable peak in the periodogram\footnote{For specificity, we designate
the star with the stronger peak in the Lomb-Scargle periodogram as the
primary -- e.g. EPIC 204060981A.  However, the data we have does not
allow us to determine which component is more massive nor which component
is more luminous.  Thus these designations should be regarded as
tentative, pending future radial velocity or high-spatial resolution
imaging studies}.  This was demonstrably true in
the Pleiades, where the dM stars with two peaks in their periodogram
formed a well-separated sequence above the single star locus  in a $V$
vs.\ $V-K_s$ CMD (see Figure 19 of Stauffer \etal\ 2016).  We discuss
binary issues more in later sections of this paper. We note that one
of the stars in Table 1, EPIC 204060981, turned out to be the first
system where both components of the binary fall into the
``scallop-shell" category.

\begin{figure}[ht]
\epsscale{0.9}
\plotone{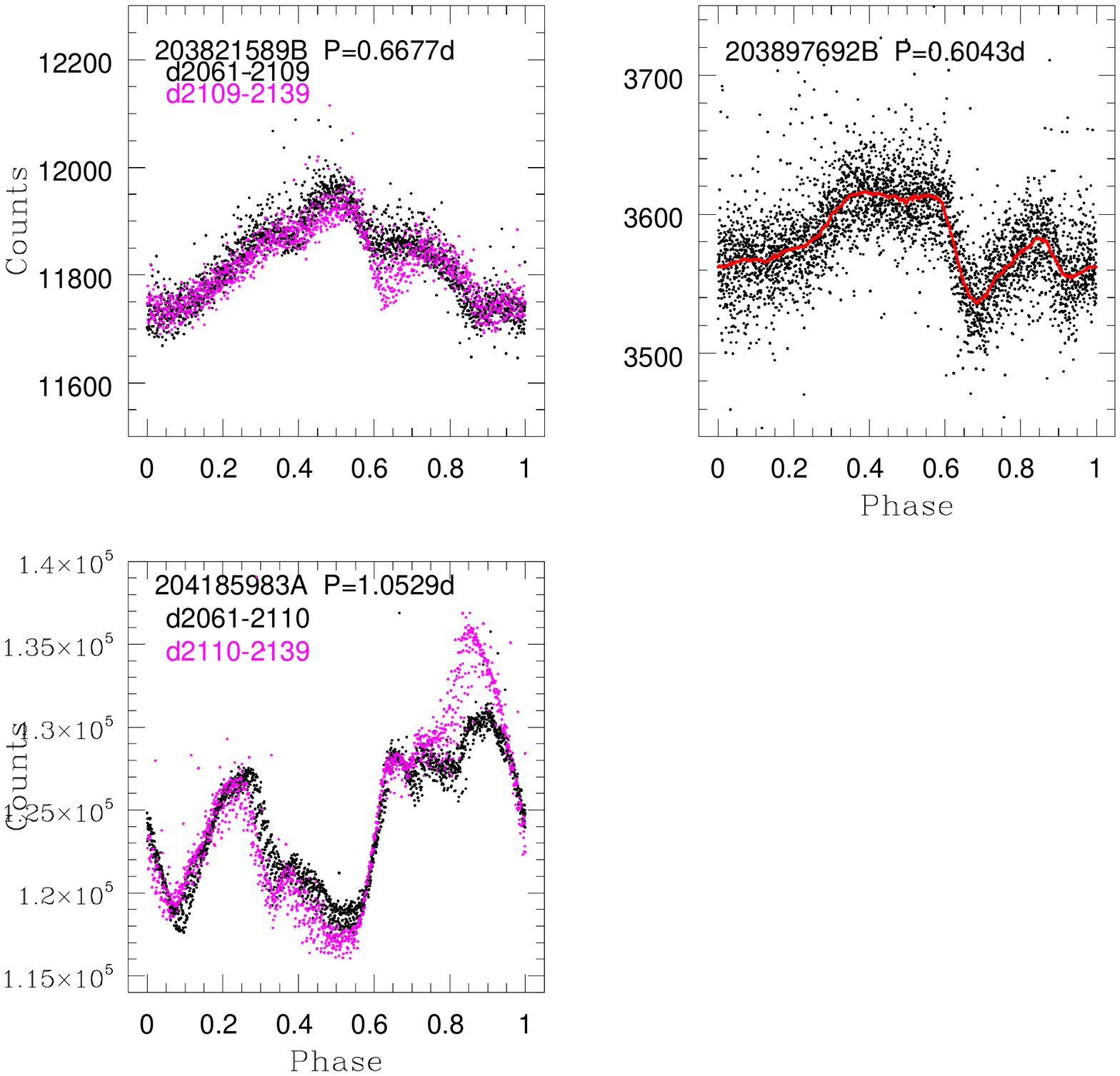}
\caption{The phased light curve  of the three stars in Table 1 that
are members of $\rho$ Oph.   See Figure 1 for a description of the colored
points and curves.  All of the phased light curves in
Figures 1 and 2 have multiple peaks and multiple local flux minima, often with
rapid changes in flux over very short intervals in phase.   These are
defining characteristics of the scallop-shell light curve class.
\label{fig:superfigure2}}
\end{figure}

\begin{figure}[ht]
\epsscale{0.9}
\plotone{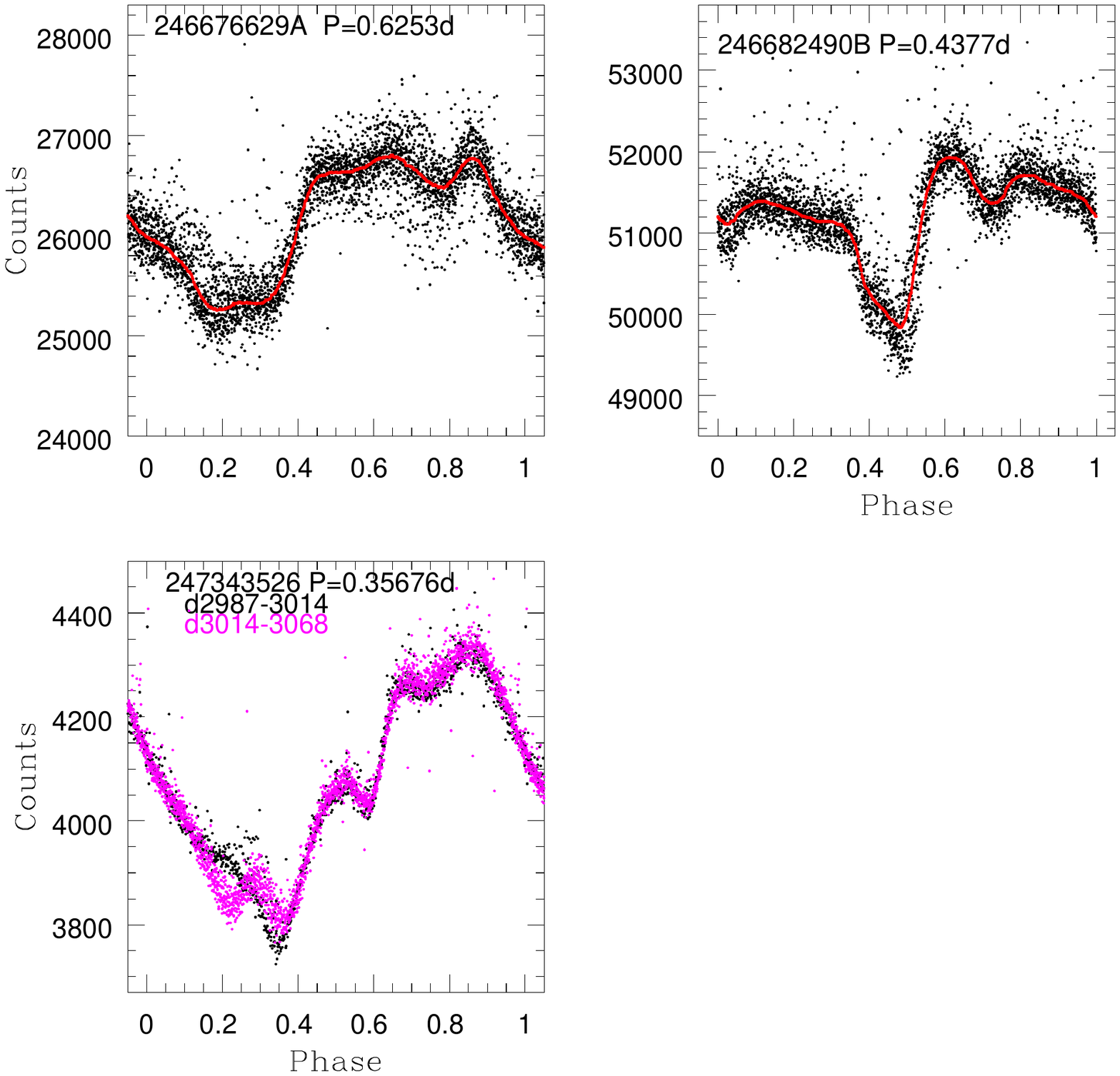}
\caption{The phased light curve  of the three stars in Table 1 that
are members of the Taurus star-forming region.
See Figure 1 for a description of the colored
points and curves.
\label{fig:superfigure3}}
\end{figure}

\section{Light Curve Morphologies of the Stars in Table 1}

Figures 1-3 show the phased light curves for all of the stars in
Table 1 except for EPIC 204099739, which we discuss separately in
\S 4.  EPIC 204060981 appears twice, because it is a binary and
both components of the binary have scallop-shell phased light curves.
Where the light curve morphology was stable or nearly
stable over the entire 78 days of the campaign, we simply show all
points in the phased light curves as black dots.  Where there was a
significant change in morphology during the campaign, we have divided
the data for the star into two or more time windows, and then plotted
the phased light curve with different color points for each time
window.  Close examination of these phased light curves shows that the
changes in light curve morphology are generally restricted to small ranges in
phase in all cases -- as was the case for the stars identified in S17.

\begin{figure}[ht]
\epsscale{0.65}
\plotone{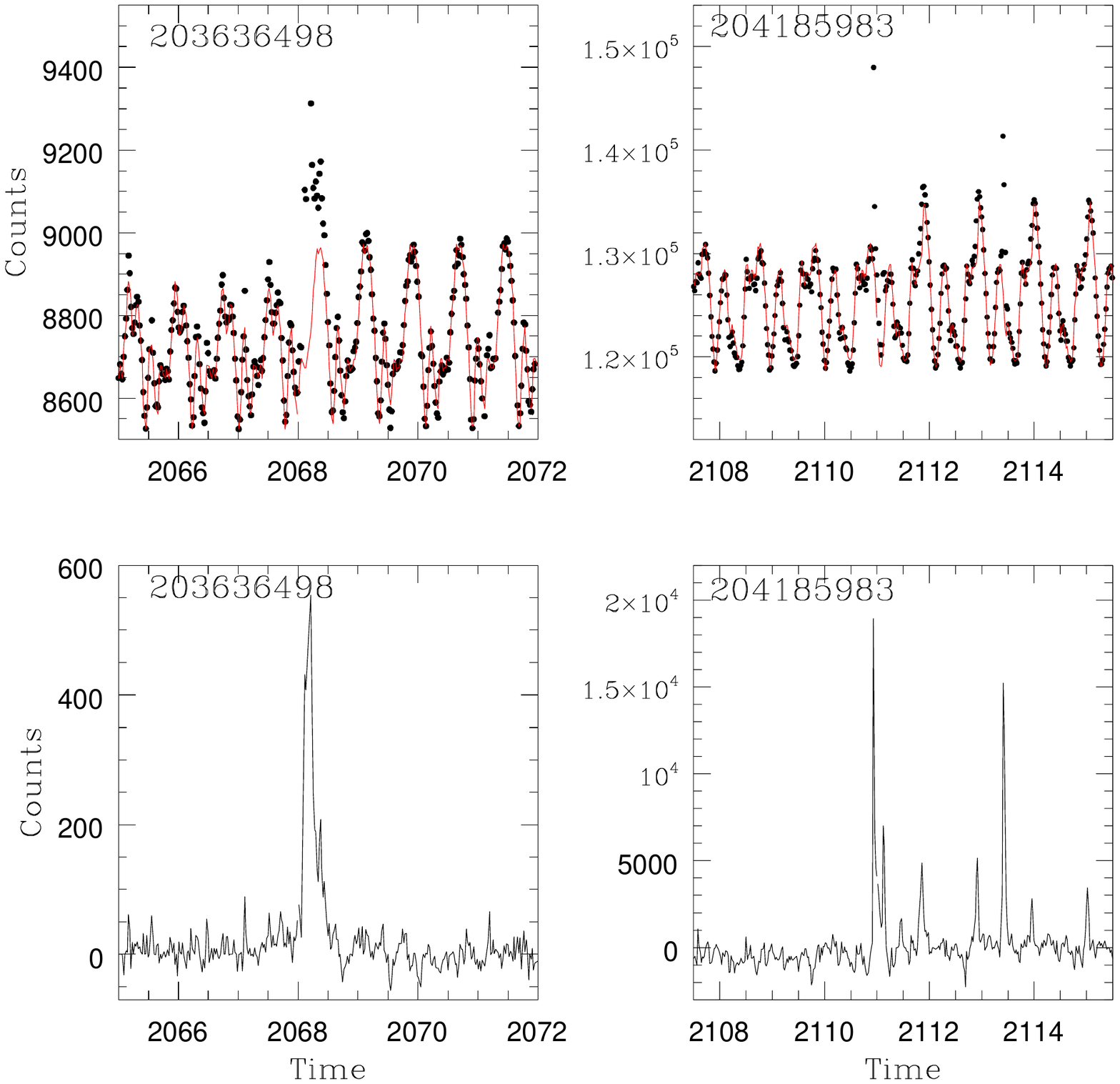}
\caption{(top) Expanded view of the light curves
for EPIC 203636498 and 204185983 around the time of the strongest
flare or flare-like event for those two stars during the K2 campaign.
Dots are the original light curve; red curves are median fits to
the phased-light curve pre- and post-flare (see text for details);
(bottom)
Plot of the same time ranges, where we have subtracted the median
fits from the data.   The residual fluxes provide a better view
of the flare-like events.   The other smaller flux peaks for
EPIC 204185983 may be artifacts of our process, except for the
peak at day 2113 which we believe is real.  Plots of the 10 day
intervals before and after the shown time interval show no other
peaks with amplitude greater than 5000 counts.
\label{fig:super_flares}}
\end{figure}

In S17, we showed that in about half the cases where the phased light
curve suddenly changed shape, the epoch where the change took place
corresponded closely to the occurrence of a flare or flare-like event
in the light curve. For the six stars in Figures 1-3 that show
changing morphology in their phased light curve, two show a
flare or flare-like event at the time of the transition.  Figure \ref{fig:super_flares}
shows an expanded view of these light curves around the time of the
change in light curve morphology.   The flare-like events occur
at day 2068.3 for EPIC 203636498 and at day 2111.0 for EPIC
204185983A\footnote{The times listed here and throughout the
remainder of the paper are in days since January 1, 2009, or as
JD - 2454833.  {\em Kepler} was launched on March 7, 2009.}.   
We have made median fits to the phased light curve
waveform for the $\sim$ 10 day period immediately prior to and after
the flare; in these plots, the median fits are shown as red curves and the original
light curves as filled dots.  We have subtracted the median fits to the
phased light curve waveform from the original light curve and plotted
the residuals in the bottom panels of Figure \ref{fig:super_flares}.
For EPIC 204185983, there are additional smaller possible flares for
day $>$ 2111.  We believe the peak at day 2113.4 may be real, but the
other small flux peaks are more likely just due to mismatches between
the median fit and the actual period-to-period shape of the light curve.
For both stars, after the
flare, the phased light curve becomes brighter at about the point in
the phased waveform when the flare occurred.   This again was what
we observed for the flares in the S17 sample.

\subsection{EPIC 204185983: The Brightest Star with a Scallop-Shell Light Curve}

In the Kepler bandpass, EPIC 204185983 is about a magnitude brighter than
any of the other stars with scallop-shell light curve morphology, which
should in principle allow us to obtain a more detailed measurement of its
light curve shape and evolution.  The third panel of Figure \ref{fig:superfigure3}
does indeed show that its light curve is very well-defined, and it appears
to have considerable morphological evolution over the K2 campaign.  For
that reason, we have examined its light curve in more detail than for
the other stars in Figures 1-3.

\begin{figure}[ht]
\epsscale{0.65}
\plotone{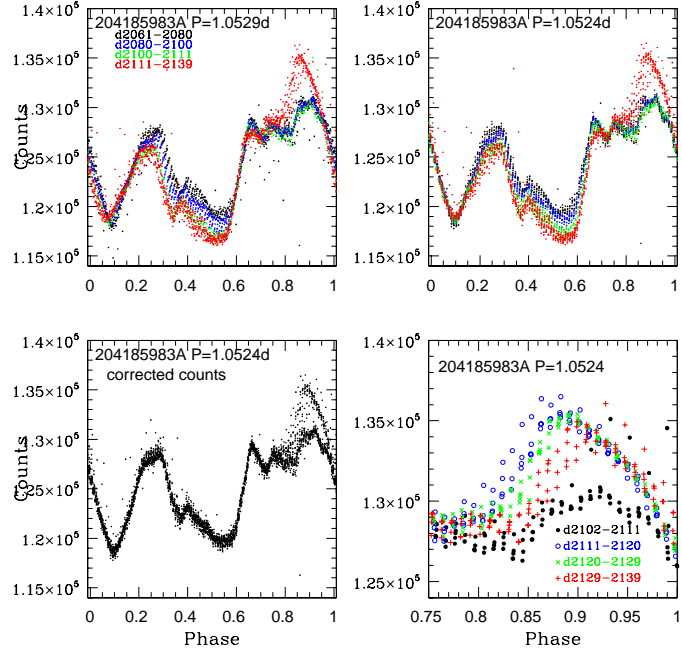}
\caption{(a) Phased light curve for EPIC 204185983A, for an assumed
period of 1.0529 days.  The four colors correspond to different time
spans during the campaign, as denoted within the panel; (b) Same as
for panel a, except now for a period of 1.0524 days;
(c) The light curve after application of a toy model to ``correct" the
light curve for a linearly decreasing flux for phases 0.2 to 0.65;
(d) Expanded view of the EPIC 204185983 light curve from just prior to
the flare illustrated in Figure 4 to the end of the campaign.  In this
case, the four colors correspond to 9 day windows beginning on day 2102.
\label{fig:super_X5983}}
\end{figure}

Figure \ref{fig:super_X5983}a repeats the plot for Figure \ref{fig:superfigure3}c,
but this time breaking the light curve up into four time windows.  The period
of 1.0529 days was chosen primarily because it yielded the least scatter near
phases 0.0, 0.2 and 0.65 where the phased light curve is changing in flux
most rapidly.  However, as illustrated in Figure \ref{fig:super_X5983}a,
between phases 0.25 and 0.6 the light curve undergoes a progressive
decrease in flux by $\sim$3\%\ and a shift in phase over the K2 campaign.
Figure \ref{fig:super_X5983}b shows that by adopting a slightly different
period of 1.0524 days, the progressive phase shift for 0.25 $< \phi <$ 0.6
is eliminated and some of the other features align better over time.

We next created a toy model with a linear decrease in flux with time at the
rate of 3\% in 80 days for phases 0.2 to 0.65, with edge tapering for
0.15 $< \phi <$ 0.2 and 0.65 $< \phi <$ 0.7.  Figure \ref{fig:super_X5983}c
shows that the toy model does a generally good job of removing the time
evolution of the light curve shape.  The last panel in Figure \ref{fig:super_X5983}
provides a closer view of the other major evolutionary change in the light
curve -- the sudden jump in flux for $\phi \sim$ 0.9 that occurred at day 2111
(as highlighted in Figure \ref{fig:super_flares}).   Figure \ref{fig:super_X5983}d
shows that after the flare, the flux excess slowly decreases with time, possibly
evolving  back to the shape it had prior to the flare.   All of these characteristics
illustrate that while there is short timescale evolution in some of the
scallop-shell light curves, there must be some underlying structure (the star's
magnetic field topology?) which remains relatively stable over the 80 day
timescale of a typical K2 campaign.

\subsection{Towards a Quantitative Criterion for Scallop-Shell Light Curves}

\begin{figure}[ht]
\epsscale{1.0}
\plottwo{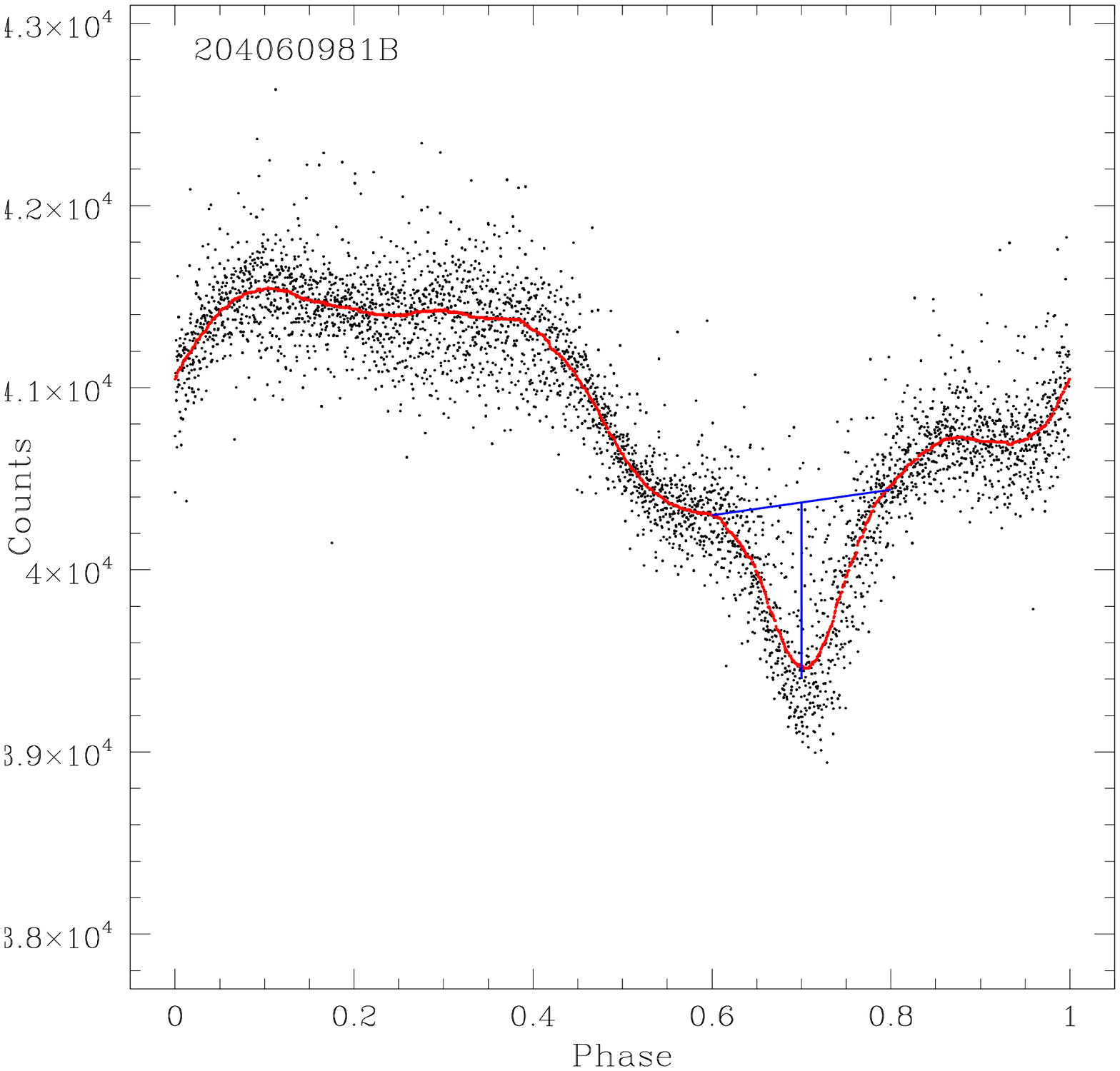}{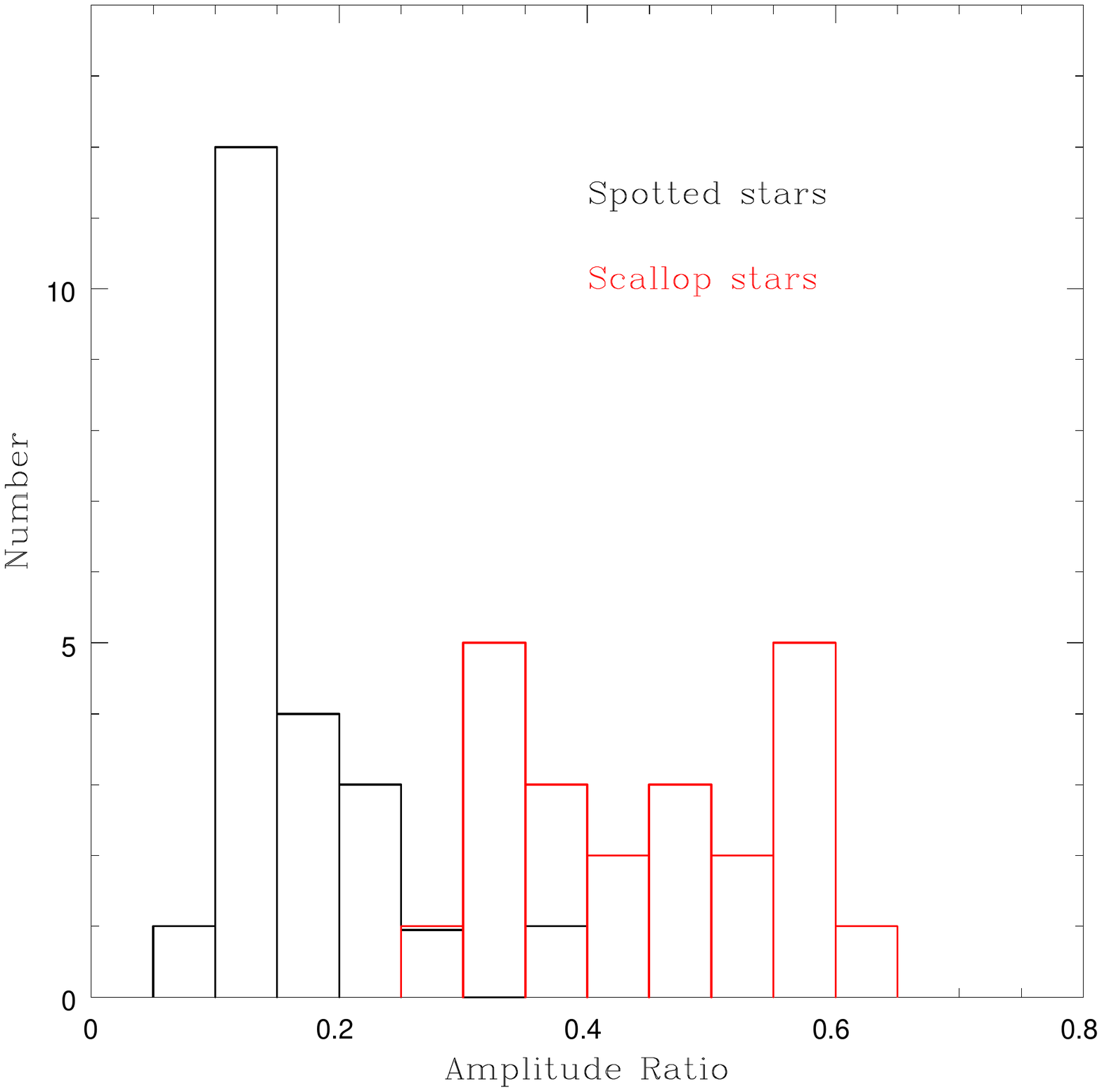}
\caption{(a) Our detrended light curve for EPIC 204060981B.  
The red curve is a median fit.   The blue line connecting between
phase 0.6 and phase 0.8 of the light curve allows us to measure a
representative flux amplitude for the flux-dip centered at phase 0.7.
The amplitude we measure is shown by the vertical blue bar.  The
parameter ``f" is then the ratio of that amplitude to the full amplitude
of the phased light curve.
(b)   Histogram of amplitude ratios (``f" values) for our scallop-shell
stars and for an equal number of Upper Sco WTT M dwarfs with spotted-star
light curves (typically sinusoidal shapes).
\label{fig:figure6}}
\end{figure}

Assignment of a given phased light curve to the scallop-shell class has
been done solely on a morphological basis (that is, not based on one or
more numerical criteria).  However, one of the morphological characteristics
is that the light curve has too much structure to be attributable to
rotational modulation of non-axisymmetrically distributed photospheric
spots.  Essentially, that means that scallop-shell light curves must have several
``humps’’ or ``dips’’ that each span less than $\sim$20\% of the full period of the
star.   To provide a partial quantitative basis for this criterion, we have
determined one measure of the light curve structure for all of the scallop-shell
stars and a comparable number of randomly selected Upper Sco
spotted stars of the same spectral type range.  The parameter we measure
is the ratio ``f" of the amplitude in counts of the most prominent ``hump"
or ``dip" in the light curve compared to the full amplitude of the entire
phased light curve.  Specifically, we slide a line segment covering 0.2
in phase across a median-fitted version of the light curve and find the
location where the separation between the line and the median-fitted curve
at the mid-point in phase is largest.   This is illustrated in
Figure \ref{fig:figure6}a for EPIC 204060981B.   For a sine-wave, there is
an analytic solution, f = 0.095.  Figure \ref{fig:figure6}b shows the histogram
of ``f" values for the scallop-shell stars and the stars with typical
spotted-star morphologies.   Most of the spotted stars have ``f" values just
slightly larger than that for a pure sine function, as expected given their
``sinusoidal" waveforms.  The scallop-shell stars have much larger
``f" values, reflective of the fact that their inclusion in the class is
based on having very structured phased light curves.  There are a few
spotted stars, however, that have ``f" values that overlap with the
scallop-shell distribution.  These are stars with ``double-dip" light
curves (McQuillen \etal\ 2013), where spots at two widely different
longitudes contribute significantly to the light curve.  The ``double-dip"
stars have fewer structures in their phased light curves than the
scallop-shells, something not captured by the ``f" parameter.

\section{EPIC 204099739: Another Scallop, a Young Eclipsing Binary, or a
Young Star with a Disintegrating Planetary Companion?}

The last star in Table 1,
EPIC 204099739, has a K2 light curve that appears superficially
common-place, but when viewed in detail is unlike any of the other K2
light curves we have examined.

\begin{figure}[ht]
\epsscale{1.0}
\plottwo{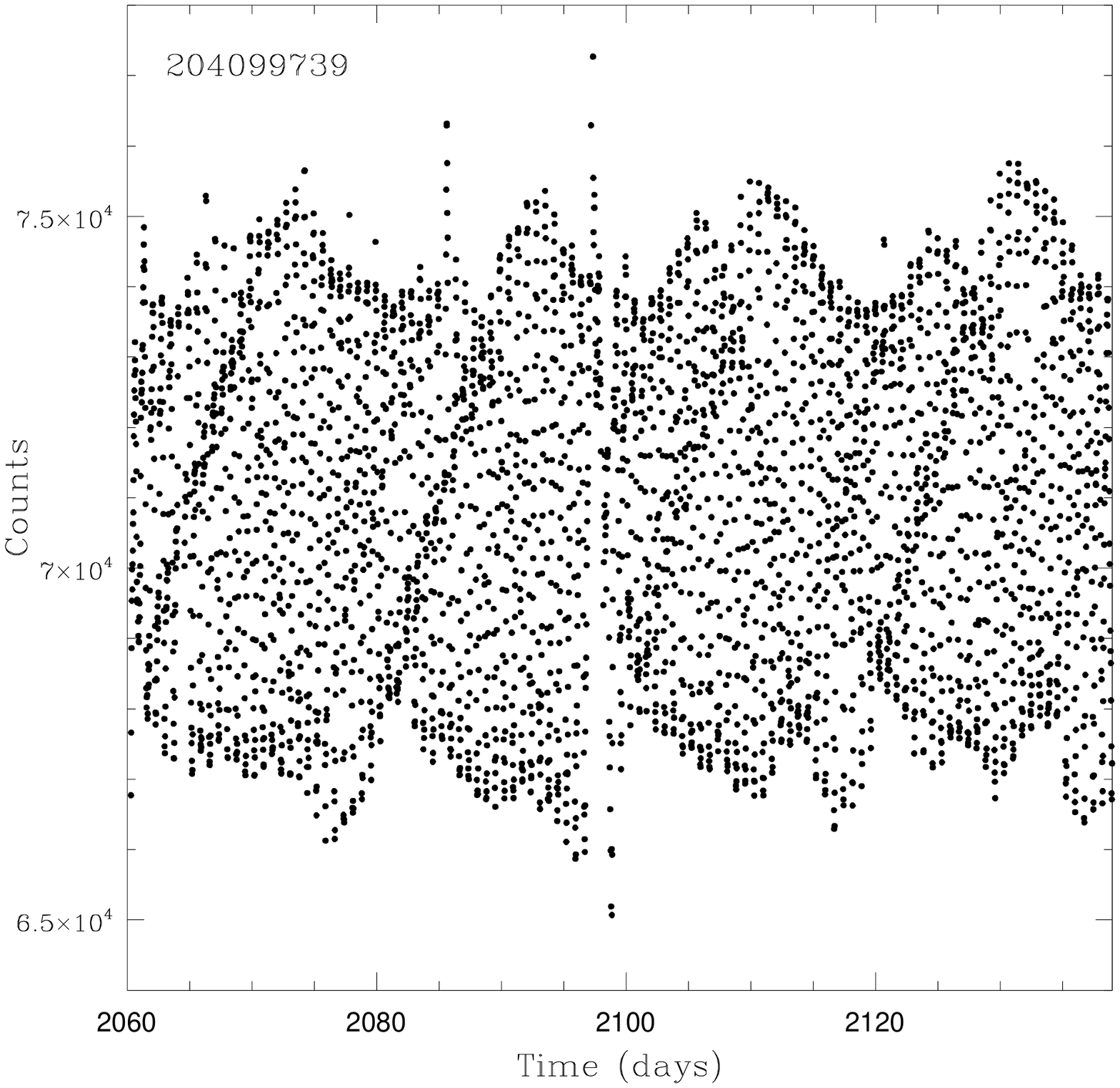}{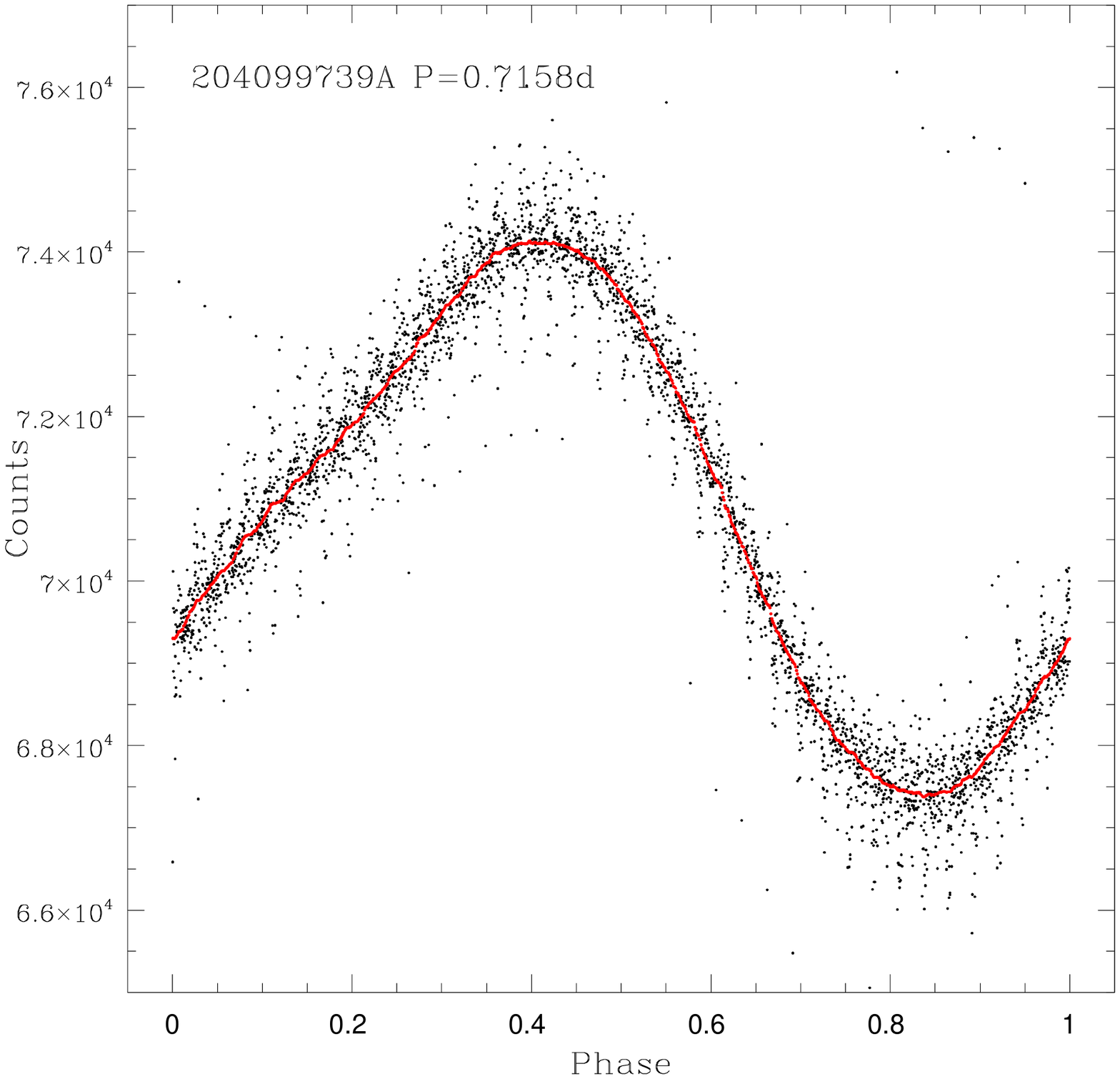}
\caption{(a) Our detrended light curve for EPIC 204099739.  The light
curve shows the typical appearance of an object with two periodic
signals with a beat cycle time of order 20 days.  A strong flare-like
event occurred around day 2098.  (b) The phased light curve for the
dominant period in the Lomb-Scargle periodogram ($P$=0.7158 days).
The light curve shows a very stable, spotted-star, sinusoidal
morphology with an  amplitude close to 10\%.   The red curve is a
median fit.
\label{fig:figure7ab}}
\end{figure}

The full, detrended K2 light curve for EPIC 204099739 is  shown in
Figure \ref{fig:figure7ab}a.  The highly structured, repetitive
pattern shown by the light curve is similar to many others in Upper
Sco, signaling that there are two periods present in the data, with
the beat cycle time being about 20 days.  The Lomb-Scargle periodogram
for EPIC 204099739 shows a very dominant $P$=0.7158 day peak with a much
weaker secondary peak corresponding to a slightly longer period of
$P$=0.743 days.   When the light curve is phased to the dominant 0.7158
day period (Figure \ref{fig:figure7ab}b), one sees a very stable,
nearly sinusoidal light curve morphology with a full amplitude near
10\%.   Light curves of this shape and amplitude are common among
young M dwarfs, and they can be modeled well with a large star-spot
at moderately high latitude seen from an inclination angle such that
at least some of the spot stays in view of the observer for the entire
rotation period.  The  stability of the light curve shape over the 78
day K2 Campaign 2 (K2C2) indicates that the spot lifetime is at least
longer than the campaign length, something that again is quite common
for young M dwarfs.

\begin{figure}[ht]
\epsscale{1.0}
\plottwo{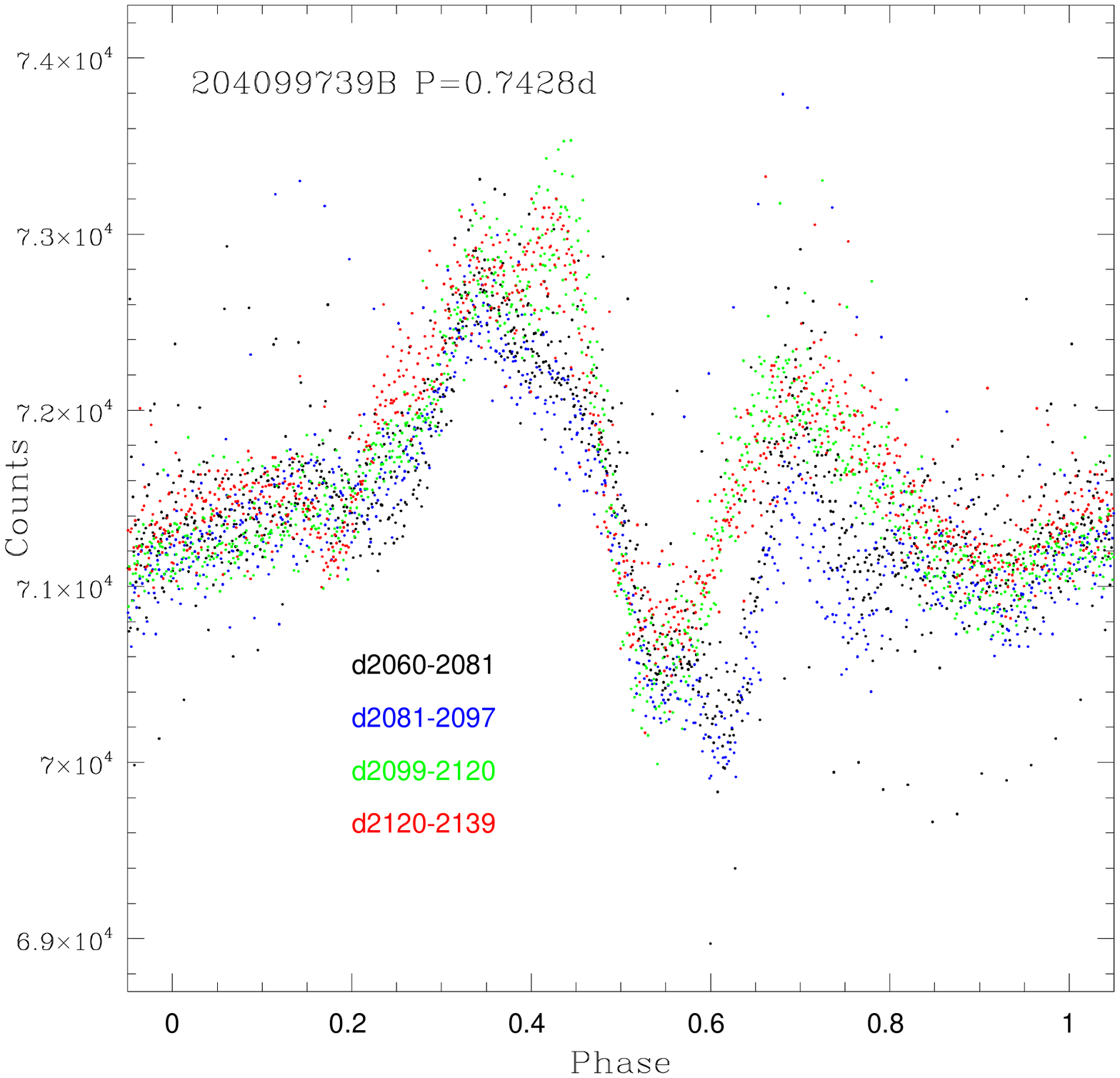}{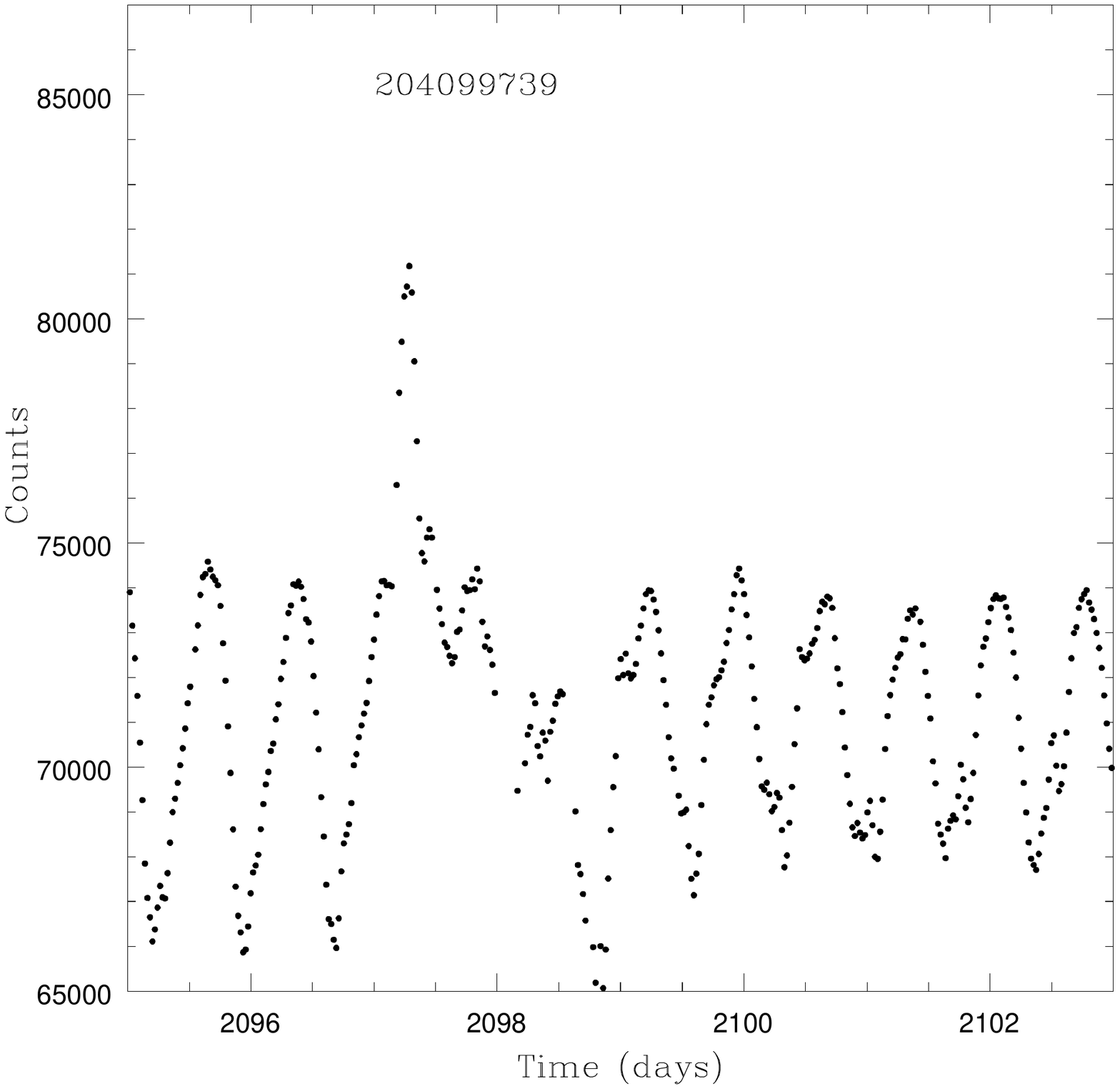}
\caption{(a) Phased light curve for EPIC 204099739
after subtracting away the light curve of the primary star, for an
adopted period of P = 0.7428 days.  The light curve shape appears to have
been stable for about the first 40 days of the campaign, then made an
abrupt small change to its morphology, followed by another 40 days
with a relatively stable shape.  The zero-points in phase and
in countrate for this light
curve are arbitrary. (b)  Expanded view of the original light curve for
EPIC 204099739, illustrating the flare-like event at about day 2097.3.
\label{fig:figure8ab}}
\end{figure}

To obtain a measure of the light curve morphology associated with the
second period, we subtracted the light curve shown in Figure
\ref{fig:figure7ab}b from  the original light curve, and then phased
the residual light curve to $P$=0.7428 days.  That phased light curve
is shown in Figure \ref{fig:figure8ab}a.  The different colored points
in the figure correspond to four approximately 20 day time windows,
from black (start of campaign), to blue, to green, to red (end of
campaign).  This phased light curve has a very unusual shape, best
characterized as two flux peaks separated by a relatively narrow flux
dip. Comparing the locus of the black and blue points to that for the
green and red points, there is apparently a small, abrupt shift in the
light curve morphology around day 2098.  Figure \ref{fig:figure8ab}b
displays the original light curve for an eight day window centered on
day 2099.  There was a relatively strong flare or flare-like event at
day 2097.3; this was by far the strongest flare for EPIC 204099739
during K2C2.

Star-spots of any kind cannot explain the light curve morphology shown
in Figure \ref{fig:figure8ab}a.   We instead consider three classes of
objects that might offer possible solutions.   Our conclusion from
this exercise is that the object associated with the 0.7428 day period
is most likely a second star whose variability arises from the
mechanism producing the scallop-shell light curves associated with the
other objects in Table 1.

\subsection{Is the P = 0.7428 day Waveform the Signature of a Young, Disintegrating Planet?}

Prototypical stars whose Kepler light curve signatures have been
attributed to disintegrating planets are KIC 12557548 (Rappaport
\etal\ 2012) and K2-22 (Sanchis-Ojeda \etal\ 2015).   Both these stars
show periodic, variable depth flux dips with  some evidence for flux
brightenings before and/or after the flux dip.  Both are short period
systems ($P$ = 16 hr and 9 hr, respectively), and both have maximum
dip depths of about 1\%.   Those characteristics bear superficial
resemblance to the light curve morphology for EPIC 204099739.
However, other aspects of the $P$ = 0.7428 day phased light curve for
EPIC 204099739 differ greatly from the disintegrating planet
prototypes.  First, the flux dip depths for KIC 12557548 and K2-22
are highly variable (by a factor of 10 in depth) even on very short
timescales, whereas for EPIC 204099739 the depths are stable except
for the single abrupt change.   Second, the amplitude of the preceding
flux peak in EPIC 204099739 is nearly 3\% and the trailing flux peak
has an amplitude of more than 1\%.  For KIC 12557548 and K2-22, those
amplitudes are all less than 0.1\%.   The extent in  phase of the dip
and flux humps for KIC 12557548 and K2-22 is also much less than what
is seen for EPIC 204099739 (Figure \ref{fig:figure8ab}a).  It seems
very unlikely that the cometary debris tail models that can explain
KIC 12557548 and K2-22 (see discussion in Sanchis-Ojeda \etal\ 2015)
could be amped up sufficiently to explain the much larger amplitude
and longer in duration flux peaks seen in EPIC 204099739.   We
therefore discard this possibility.

\subsection{Is the P = 0.7428 day Waveform the Signature of a Heartbeat Binary?}

Another recent discovery from the Kepler mission light curves are the
so-called heartbeat stars (Thompson \etal\ 2012; Smullen \& Kobulnicky
2015; Zimmerman \etal\ 2017).   These stars can show complex phased
light curve waveforms, including sometimes having two flux peaks
straddling a relatively narrow flux dip.  The amplitudes of both the
peaks and the dips can at least approach 1\%.  These systems have been
found to have relatively short periods and  moderate to high
eccentricities; their photometric variability is primarily ascribed to
tidal distortions that are strongest near periastron.   Models of such
systems (e.g., Figure 5 of Thompson \etal\ 2012) can yield flux peak
and dip signatures that closely resemble those seen for EPIC
204099739 in shape and phase width (the light curve signature depends
sensitively on eccentricity, observer inclination angle and periastron
view angle). However, we again believe that in detail, EPIC
204099739's light curve is quite unlikely to result from this
mechanism.   The decisive argument  is that, at least over the
timescale of the K2 campaign, the tidally induced photometric
variability should be essentially fixed in morphology.  The observed
abrupt change in light curve shape at day 2097 seems completely
incompatible with this mechanism.

\subsection{Is the P = 0.7428 day Waveform the Signature of a Companion Star with a
    Scallop-Shell Light Curve?}

We did not originally consider EPIC 204099739 as an obvious member of
the scallop class. The other members of the class have various
combinations of flux peaks and flux dips, but none that resembled two
flux peaks separated by a flux dip.   However, in essentially all
other aspects, EPIC 204099739 does seem compatible with that
category:
\begin{itemize}
\item  It has a very short period, as do all other members of the class.
\item  It has a photometric amplitude of at least 4\%,
which is typical of the class (but see below).  
\item It shows an abrupt change in phased light curve morphology which
occurs at the time of a flare or  flare-like event, one of the most
striking characteristics of the class.   
\item It is a young M dwarf with no evidence of on-going accretion nor of
a primordial dust disk.
\end{itemize}

If the P = 0.7428 day period corresponds to a scallop-shell waveform,
then to fit our other assumptions EPIC 204099739 must be a binary
star -- with one star corresponding to the $P$ = 0.7158 day sinusoidal
variation and the other star having a rotation period of $P$ = 0.7428
days.  We have little knowledge of the orbital period of the system;
our only insight comes from our two spectra of the system, both of which show single
lines and radial velocities that  differ at most slightly from that
expected for an Upper Sco member.  This offers some evidence against
this being a close binary, but more spectra or AO (adaptive optics)  imaging are needed
to confirm.  We also have no certain knowledge of the relative
brightness of the two stars, but we can place constraints on their
relative contribution to the K2 light curve.   The amplitude of
variability for the $P$ = 0.7158 day waveform is about 10\%.  That is
actually a lower limit since some fraction of the system light comes
from the second star.   Very few young dM stars have photometric
variability amplitudes exceeding 20\% at R band.   Therefore, it is
unlikely that the second star contributes more than 50\% of the light
to the K2 light curve.  If the second star did contribute 50\% of the
light, then the true amplitude of variability for the scallop-shell
component would be about 8\%, which is close to the maximum seen for
any of the scallop-shell stars presented in this paper or in S17.

\section{Physical Properties of the Stars having Scallop Shell Light Curves}

The preceding sections have focused entirely on the K2 light curves
for the scallop-shell stars.  The light curve morphologies, periods,
amplitudes, and evolution with time presumably place strong constraints
on possible physical mechanisms that could explain the photometric
variability of these stars.  Now that we have a sample of $\sim$20
of these stars, it is useful to see how their general (non-time-domain)
properties compare to other stars of the same mass and age, as that too
could help constrain possible physical mechanisms.

\begin{figure}[ht]
\epsscale{1.0}
\plotone{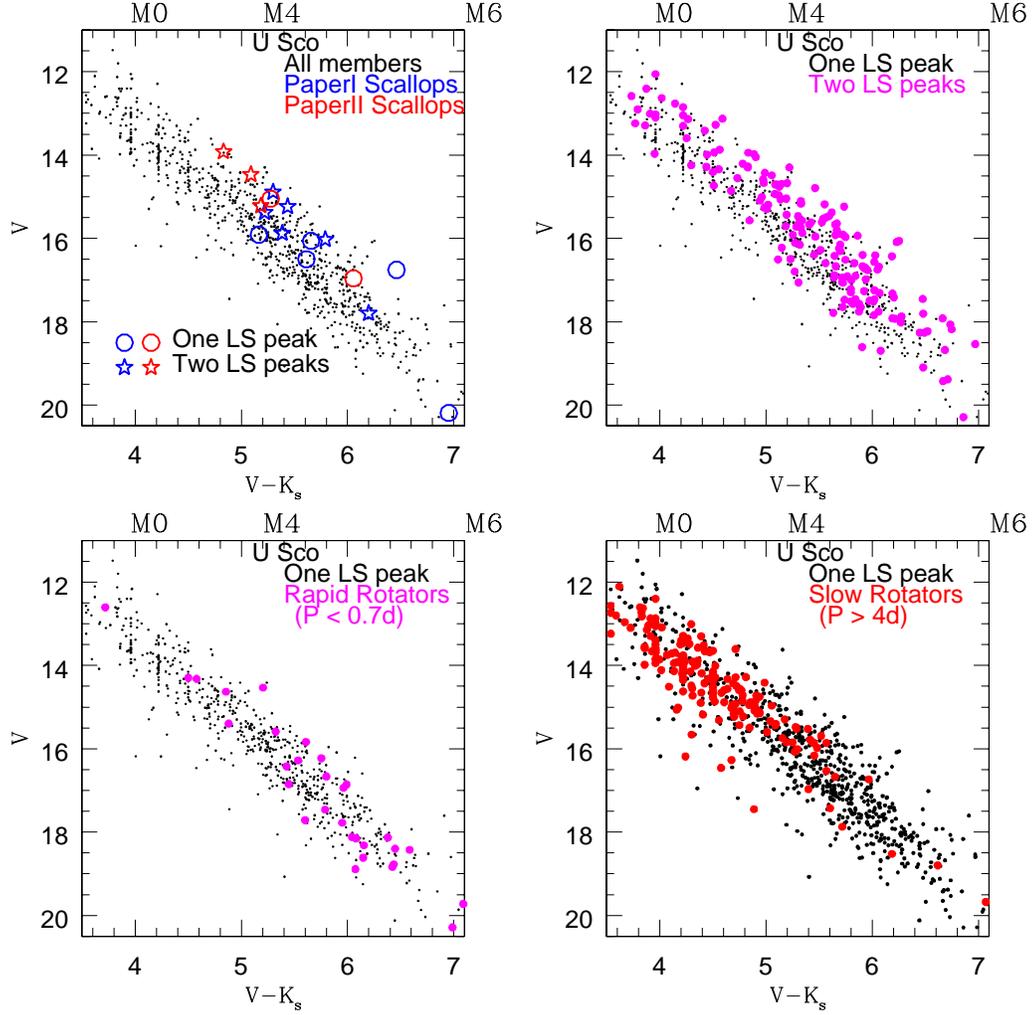}
\caption{(a)  CMD for Upper Sco members, with the stars having
scallop-shell light curves highlighted.   The scallop-shells are shown
as star symbols if they are in binary systems with two peaks in their
periodograms or as open circles if we detect only one peak. Black dots
are all other Upper Sco members with K2 periods. (b)  CMD for Upper
Sco, this time highlighting all stars with two periodogram peaks.  The
stars with two periods are systematically  displaced towards the
bright half of the Upper Sco locus as is  expected for binaries with
relatively large mass ratios. (c) CMD highlighting the most rapidly
rotating dM Upper Sco members. (d) CMD highlighting the slowly
rotating dMs.   All plots utilize dereddened photometry based on A$_V$'s
estimated from a $J - H$ vs. $H - K$ color-color diagram.
\label{fig:cmds_scallops}}
\end{figure}

Figure \ref{fig:cmds_scallops} shows several versions of a $V_o$ vs.\
$(V-K_s)_o$ CMD for Upper Sco.   The first panel highlights the stars
with scallop-shell light curves.  It demonstrates that these stars
are, on average, towards the bright side of the locus, suggesting that
they are either younger or more often in binary systems.  The next
panel specifically highlights stars we detect as binaries (via two
peaks in the K2 periodogram), which confirms that these stars are also
displaced to the bright side of the Upper Sco locus, as expected.
Panel (c) highlights the most rapidly rotating dM members.  They too
appear, on average, to be younger and/or weighted towards being
binaries.   Finally, the last panel shows a similar plot for the more
slowly rotating dM Upper Sco members.  Other than not sampling well
the lowest mass (and hence faster rotating) part of the population,
the slow rotators do not show any obvious displacement relative to the
mean cluster locus in the CMD.  

The simplest synthesis of what is shown in these CMDs is that at Upper
Sco age, M dwarf members of binary systems tend to be rapid rotators
compared to their single (or low-q binary) cousins, where $q = M_1/M_2$.  
We had reached a
similar conclusion previously for the M dwarfs observed with K2 in the
Pleiades (Stauffer \etal\ 2017). That about half of the stars
with scallop-shell light curves are in binaries is likely just a product
of two correlations: (a) the fact that the physical mechanism producing
the scallop-shell light curves is likely linked to very rapid rotation; and
(b) the fact that components of binaries tend to be the most rapidly rotating
M dwarfs in Upper Sco.

\begin{figure}[ht]
\epsscale{1.0}
\plotone{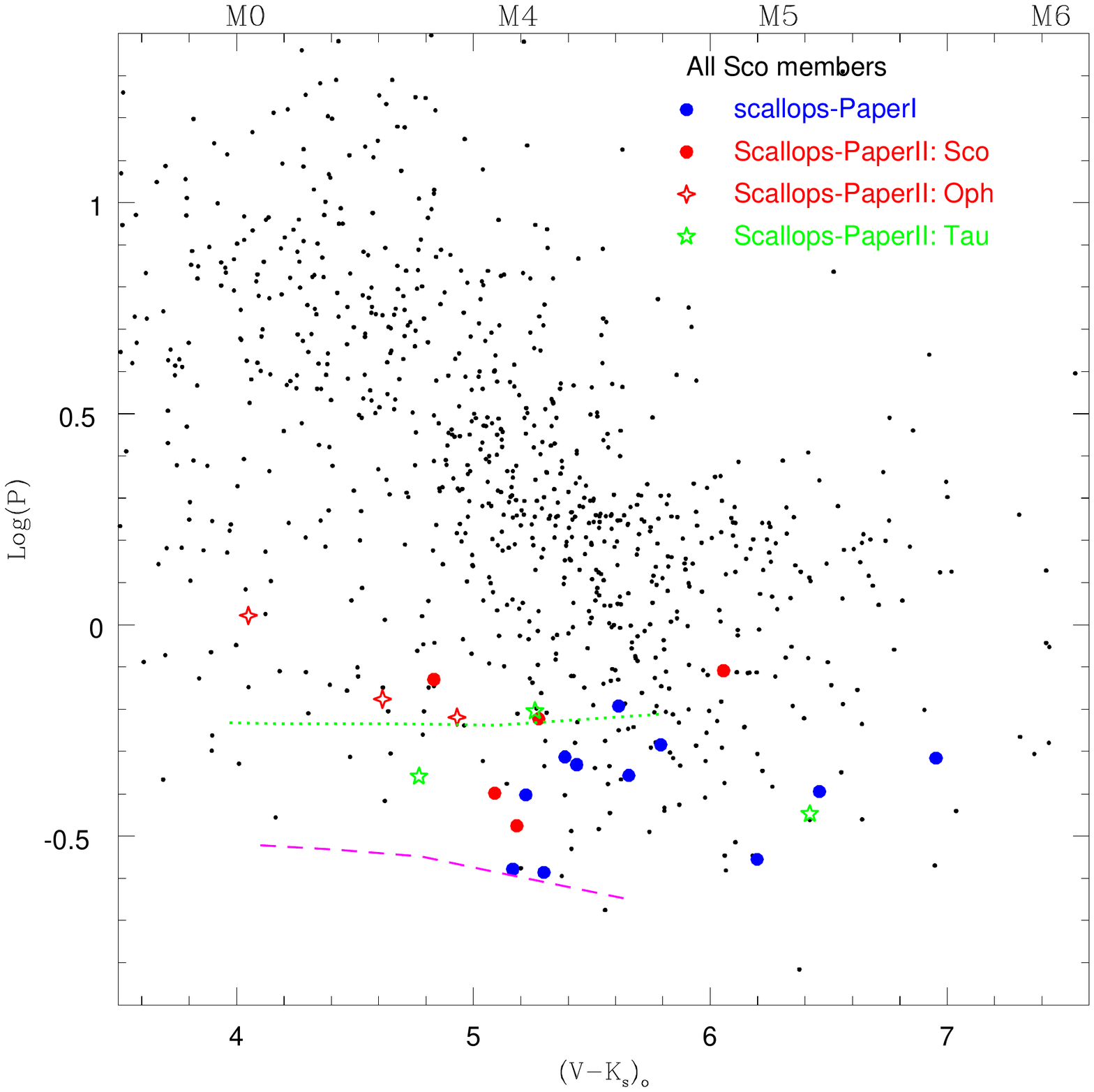}
\caption{ Period-color diagram for all the stars with scallop-shell
light curves identified to-date in Upper Sco, $\rho$ Oph and Taurus.
Black dots are all Upper Sco M dwarfs with K2 rotation periods.
The long-dashed magenta line is an estimate of the breakup period assuming an
age of 8 Myr; the short-dashed green line is the corresponding curve
for 2 Myr.   All of the stars with scallop-shell light curves have
very short periods.  The  three $\rho$ Oph scallop-shell stars are, on
average, earlier type and have longer periods relative to their Upper Sco
peers.  
\label{fig:figure_scallops_rotation}}
\end{figure}

Direct demonstration of the relatively rapid rotation of the stars with
scallop-shell light curves is provided in Figure
\ref{fig:figure_scallops_rotation}.  The scallop-shell stars clearly
segregate to the most rapidly rotating portion of the diagram.  The
most slowly rotating stars with scallop-shell light curves are those
in $\rho$ Oph -- where the much younger age means that even dM stars
with 1 day periods are rotating relatively close to breakup.  The Taurus
scallop-shell stars have periods intermediate between the $\rho$ Oph
average and the Upper Sco average, indicating perhaps that Taurus is
intermediate in age between those two clusters.
We have used Maeder's (2009) formula for the breakup period:
\begin{math}
P_{\rm breakup}^{2} = (27/8) \times (4\pi^2 R_p^3/GM_*)
\end{math}
where $R_p$ is the star's polar radius when rotating at breakup
and we have used the BHAC15 (Baraffe \etal\ 2015) models
to provide radii for a given mass and age.  The curve representing
the breakup period at 8 Myr makes a reasonable lower bound to the
Upper Sco period distribution, lending some credence to this prediction.
However, the curve for the breakup period at 2 Myr is perhaps
surprising.  If valid, it would suggest that the $\rho$ Oph stars
are of order 2 Myr old, and that at least the WTTs in Taurus
have an age of 3-4 Myr (as recently suggested by Rees \etal\ 2016 for
other reasons), rather than the more traditional age of 1-2 Myr.  
However, the radii we use to estimate the
breakup period are uncertain for several reasons and this may allow
the curves to shift up or down by a significant amount.  We note for
now simply that plots such as this might eventually be useable as
a test of the theoretical evolutionary models or perhaps as an
additional means to estimate ages of PMS clusters.

The fraction of the Upper Sco M dwarfs that are scallops is a strong
function of rotation rate: (a) none of the Upper Sco stars with $P >$ 1.0 days
are scallops; (b) 9\% (16 of 179) of those with $P <$ 1 day are
scallops; (c) 15\% (15 of 101) of those with $P <$ 0.7 days are
scallops; and (d) 22\% (12 of 55) of those with $P <$ 0.5 days are
scallops.  Models to explain the scallop-shell light curves probably require
there to be a torus of gas and dust at the Keplerian co-rotation
radius and that our line-of-sight pass through that torus (Townsend \&
Owocki 2005; Stauffer \etal\ 2017; Farihi, von Hippel \& Pringle 2017;
D'Angelo \etal\ 2017).   Therefore, the percentage of stars with
scallop-shell light curve morphologies are lower
limits to the fraction of stars having these tori; for plausible
torus scale heights, the majority of the rapidly rotating young dM
stars may be so affected.  Also, several more of the rapidly rotating dMs
have marginally detected light curve features which we suspect arise from
the same processes as for the stars in Table 1 (we provide a complete
list of all the possible members of the scallop-shell class in
Rebull \etal\ 2018); we have concentrated on a highly reliable sample
of scallop-shell stars in this paper and in S17 in order to allow us to
best illustrate their light curve morphologies.

\section{Summary and Next Steps}

Using the data presented in this paper and in S17, we have identified
a previously unknown type of variable star.  Based on our current sample
of 23 such stars, the ``scallop-shell"
variability class has the following characteristics:
\begin{itemize}

\item  They all have phased light curves with multiple humps and
 dips and too much small-scale structure to be accounted for by
 rotational modulation of photospheric spots;

\item  They all have very short periods, ranging from 0.26 days to 1.05 days;

\item  With one exception, they all have mid-M spectral types, ranging from
  M3 to M6.  The one exception has spectral type M1.5.  None of these
  stars show a detected IR excess;

\item All of the members of the class have ages believed to be $<$ 10 Myr.
  A similar-sized sample of rapidly rotating M dwarfs with age $>$ 100 Myr
  show no examples of stars with this light curve morphology;

\item The light curve morphology is most commonly stable over timescales
  corresponding to a K2 campaign duration ($\sim$75-80 days);  

\item  For the minority of the members that do show light curve morphology
 changes, the change occurs abruptly and usually only affects $\sim$20\%
 of the light curve (in phase).  In about half the cases where there is
 a jump in light-curve morphology, a flare-like event is present at the time
 of the morphology change;

\item  The light curve amplitudes in the Kepler bandpass range from a couple
  per cent to more than 15\%;

\item  Among the most rapidly rotating M dwarfs in  Upper Sco, more than 20\%
  of the stars have scallop-shell light curves.  Assuming a physical model
  where the photometric variability of these stars depends on our line of sight
  intersecting a torus of gas and dust orbiting the star, half or more of the
  most rapidly rotating M dwarfs in Upper Sco may be subject to this physical
  process.

\end{itemize}

We view our current paper and our previous paper (S17) primarily
as discovery papers.   By that, we mean our goal is primarily to make
the community of young star researchers aware that there are two
dozen young PMS stars in three of the nearest sites of active star
formation that have striking photometric variability not well-explained by any
of the standard mechanisms.  We have attempted to provide as clear a description
of the empirical properties of the light curve morphologies of these
stars as possible.  However, there is very
much left to do.   The most pressing need
is to derive a detailed physical model, though some progress
has been recently made on that front (D'Angelo \etal 2017; Farihi \etal 2017).
High spatial-resolution imaging and multi-epoch radial velocity data for the
scallop-shell stars in binary systems to determine their orbital separation
would help identify any direct physical link between binarity and scallop-shell
photometric variability.  We doubt if there is such a link, but the large fraction
of the scallop-shell stars that are in binary systems argues in favor of making
this test.  Simultaneous, multi-wavelength synoptic monitoring of a few of the
stars with the largest K2 photometric variability to determine the dependence of
the variability amplitude with wavelength would directly constrain the physical
mechanism driving the variability.   New synoptic photometry at any epoch would help
determine the temporal stability of the light curve morphologies.  A few of the
scallop-shell stars have poorly determined kinematics and hence their membership
in Upper Sco, $\rho$\ Oph or Taurus is uncertain - accurate radial velocities
plus improved distances and proper motions (which should soon become available
from Gaia DR2) will help in this regard.

If the unusual light curve morphologies we have found were simply
curiosities, the time and effort needed to explain their properties
might not be justifiable.   However, given the current emphasis of targeting
M dwarfs -- and in particular late M dwarfs -- in the search for the
nearest rocky, habitable zone planets, our Upper Sco and Taurus stars take on
more significance.   A large fraction of the low mass M dwarfs in
Upper Sco may have the gas (and dust) tori that we believe give rise
to the scallop-shell light
curves (D'Angelo \etal 2017).  Understanding
the formation and evolution of these tori may well have consequences
for understanding the angular momentum evolution of these stars.  For
those reasons, we hope that other teams will work to unlock the secrets
of these stars and determine their importance to understanding the
early evolution of M dwarfs.

\begin{acknowledgements}

Some of the data presented in this paper were obtained from the
Mikulski Archive for Space Telescopes (MAST). Support for MAST for
non-HST data is provided by the NASA Office of Space Science via grant
NNX09AF08G and by other grants and contracts. This paper includes data
collected by the Kepler mission. Funding for the Kepler mission is
provided by the NASA Science Mission directorate. This research has
made use of the NASA/IPAC Infrared Science Archive (IRSA), which is
operated by the Jet Propulsion Laboratory, California Institute of
Technology, under contract with the National Aeronautics and Space
Administration. This research has made use of NASA's Astrophysics Data
System (ADS) Abstract Service, and of the SIMBAD database, operated at
CDS, Strasbourg, France. This research has made use of data products
from the Two Micron All-Sky Survey (2MASS), which is a joint project
of the University of Massachusetts and the Infrared Processing and
Analysis Center, funded by the National Aeronautics and Space
Administration and the National Science Foundation. The 2MASS data are
served by the NASA/IPAC Infrared Science Archive, which is operated by
the Jet Propulsion Laboratory, California Institute of Technology,
under contract with the National Aeronautics and Space Administration.
This publication makes use of data products from the Wide-field
Infrared Survey Explorer, which is a joint project of the University
of California, Los Angeles, and the Jet Propulsion
Laboratory/California Institute of Technology, funded by the National
Aeronautics and Space Administration.
\end{acknowledgements}

\facilities{K2, Exoplanet Archive, IRSA, 2mass, Keck:I (HIRES), Hale (DBSP),
   SOAR (Goodman), Shane (Kast Double spectrograph)}

\clearpage

\appendix

\section{New Spectroscopy}

We have obtained high resolution spectra for EPIC 203636498 and 204099739
using the Keck HIRES spectrograph (Vogt \etal\ 1994).   These
spectra cover the wavelength range roughly 4800 to 9200 \AA, at a
spectral resolution of about R = 45,000.  We have obtained
intermediate resolution spectra for EPIC 204060981, 205267399, 
204185983 and 204099739 using
the Palomar 5m Double Spectrograph (Oke \& Gunn 1982), with one star being in
common with the HIRES data (EPIC 204099739). These spectra cover the
wavelength range roughly 3800 to 7400 \AA, at a spectral resolution of
about R = 3000 for the blue-arm and R = 6000 for the red arm.  We have
obtained low resolution spectra for two of the Taurus members, EPIC 246676629 and
246682490, with the 
Lick 3m KAST spectrograph. The spectral resolution for these spectra
is R $\sim$ 1000, and the wavelength range is 5600-7750 \AA.
Finally, we have obtained an additional low resolution spectrum 
for EPIC 246676629 using the SOAR telescope and Goodman spectrograph
(Clemens \etal\ 2004).  The
resolution in this case is R $\sim$ 850 and the wavelength coverage
is 5000-9000 \AA.

We measured equivalent widths for H$\alpha$ and lithium using the
SPLOT routine in IRAF for all of these spectra.  All of the stars have
H$\alpha$  equivalent widths in the range -2.5 to -10, in accord with
expectations for WTTS of a few Myr age.  Where our spectra have sufficient
S/N, all of them but one show lithium absorption equivalent widths
also in the range expected for early to mid M dwarfs of this age.  
The one exception is EPIC 205267399, which has a good S/N DBSP
spectrum but a limit of its lithium equivalent  width of $<$ 0.15
\AA.   Everything else suggests this star is a member of Upper Sco, so
we retain it in our sample.   The relatively low spectral resolution for
the KAST and Goodman spectra is 
not optimal for measurement of the lithium feature, but 
we believe it is present in both EPIC 246676629 (both spectra) and
EPIC 246682490.  The lithium equivalent widths for these stars are uncertain
at probably the 50\% level, which we indicate in Table 1 by adding a
colon after the value we have measured.

We have also used these spectra to determine
spectral types based
on the strength of the TiO bandheads in the 6500-7500 \AA\ range (Stauffer \etal\
1979; Preibisch \etal\ 2001). The estimated spectral types are listed in Table
1.

\begin{figure}[ht]
\epsscale{0.65}
\plotone{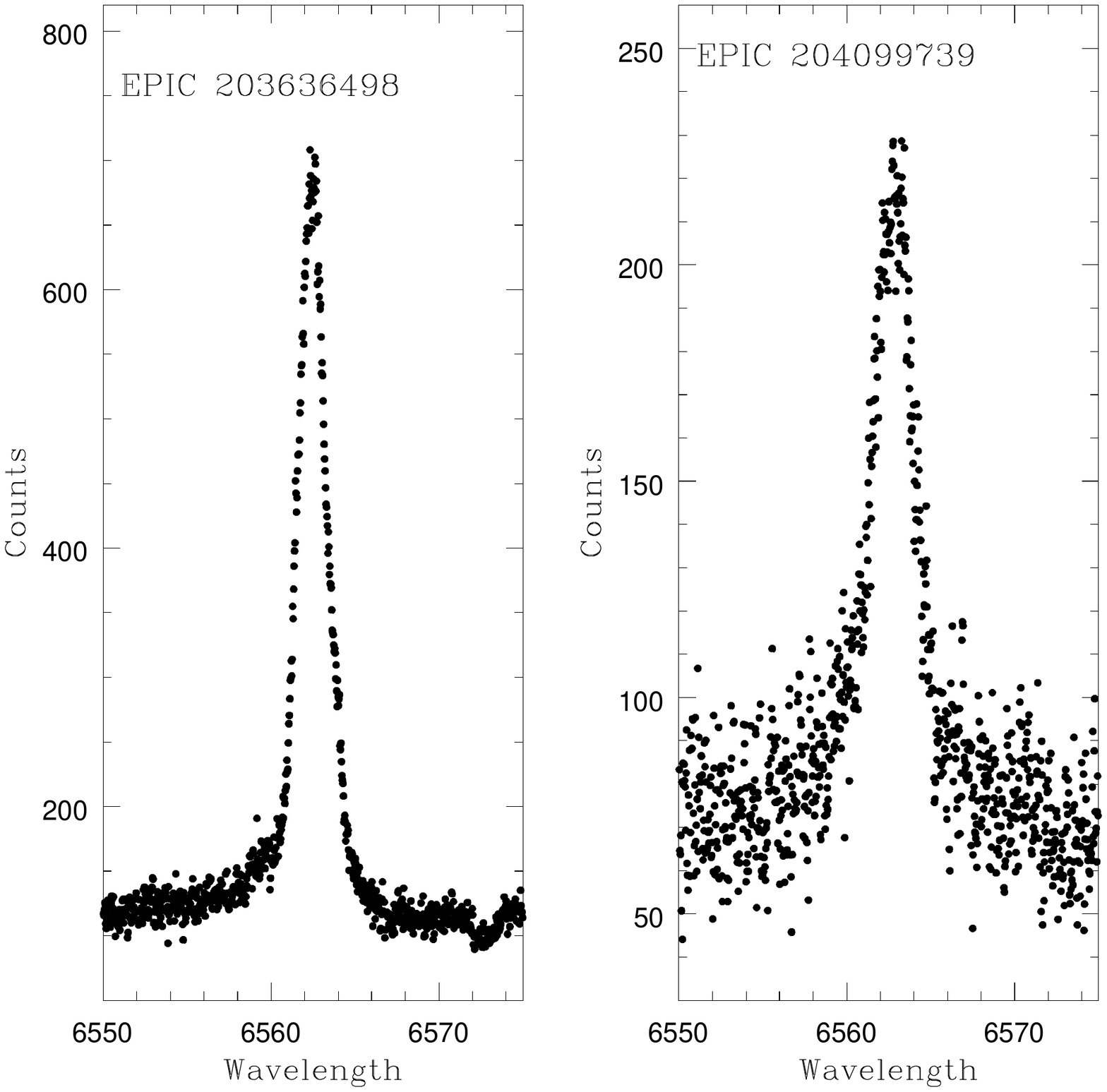}
\caption{H$\alpha$\ emission line profiles from our HIRES spectra.
\label{fig:hires_spectra}}
\end{figure}

\begin{figure}[ht]
\epsscale{0.65}
\plotone{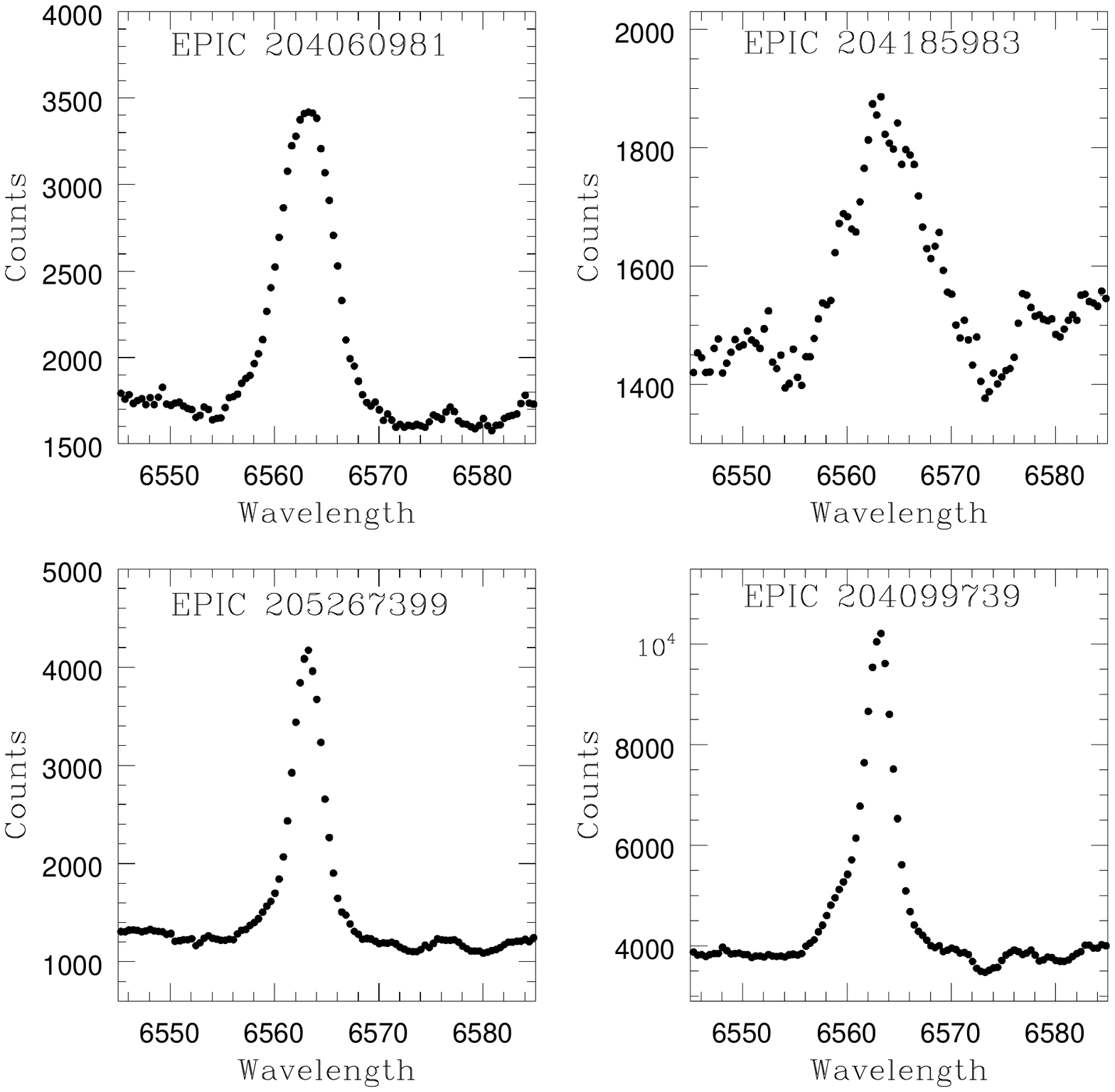}
\caption{H$\alpha$\ emission line profiles from our DBSP spectra.
\label{fig:dbsp_spectra}}
\end{figure}

\begin{figure}[ht]
\epsscale{0.65}
\plotone{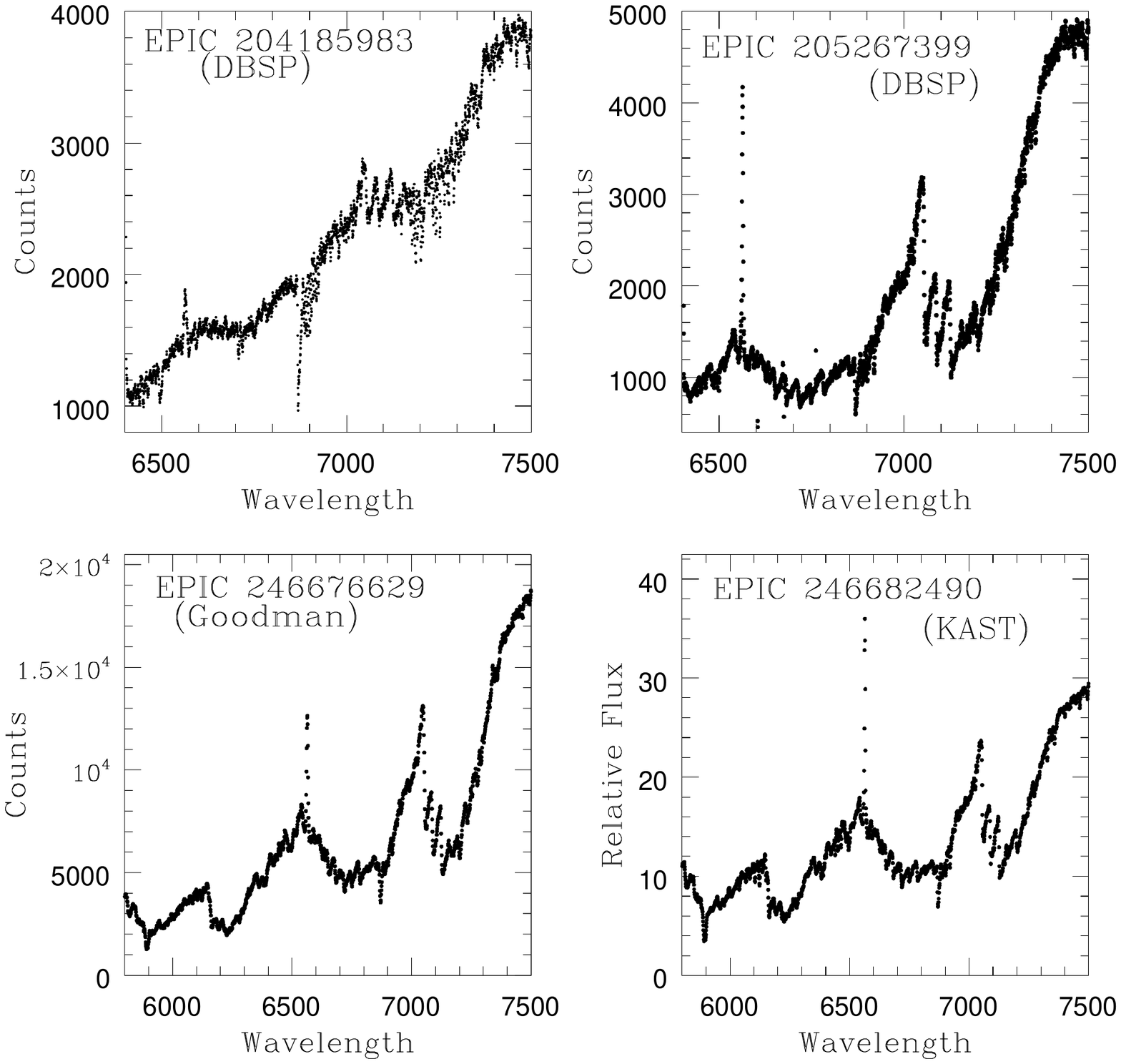}
\caption{Red-optical spectra from our DBSP, Goodman and KAST spectra for four
of our stars.  Their inferred spectral types range from M1.5 to M5.
Note that the KAST spectra have been flux calibrated, whereas the Goodman and 
DBSP spectra are simply calibrated to counts per pixel.  EPIC 204185983 is
both significantly earlier spectral type and significantly much more reddened
than the other three stars.  EPIC 204185983 is a member of $\rho$\ Oph,
EPIC 205267399 is a member of Upper Sco, while the stars with Goodman and  KAST spectra
are both members of Taurus.
\label{fig:full_spectra}}
\end{figure}

Figure \ref{fig:hires_spectra} show the H$\alpha$ profiles from our
HIRES spectra; Figure \ref{fig:dbsp_spectra} show the H$\alpha$
profiles from our DBSP spectra.  The H$\alpha$\ emission lines have
approximately Gaussian shapes, as expected for rapidly rotating dMe
stars.   A few of the profiles have possible weak, high velocity wings,
but higher S/N spectra would be needed to confirm this.   
Figure \ref{fig:full_spectra} shows
the red-optical wavelength range for several of the stars observed with
DBSP, Goodman and KAST to illustrate the data available for spectral typing.

\section{Spectral Energy Distributions}

\begin{figure}[ht]
\epsscale{0.65}
\plotone{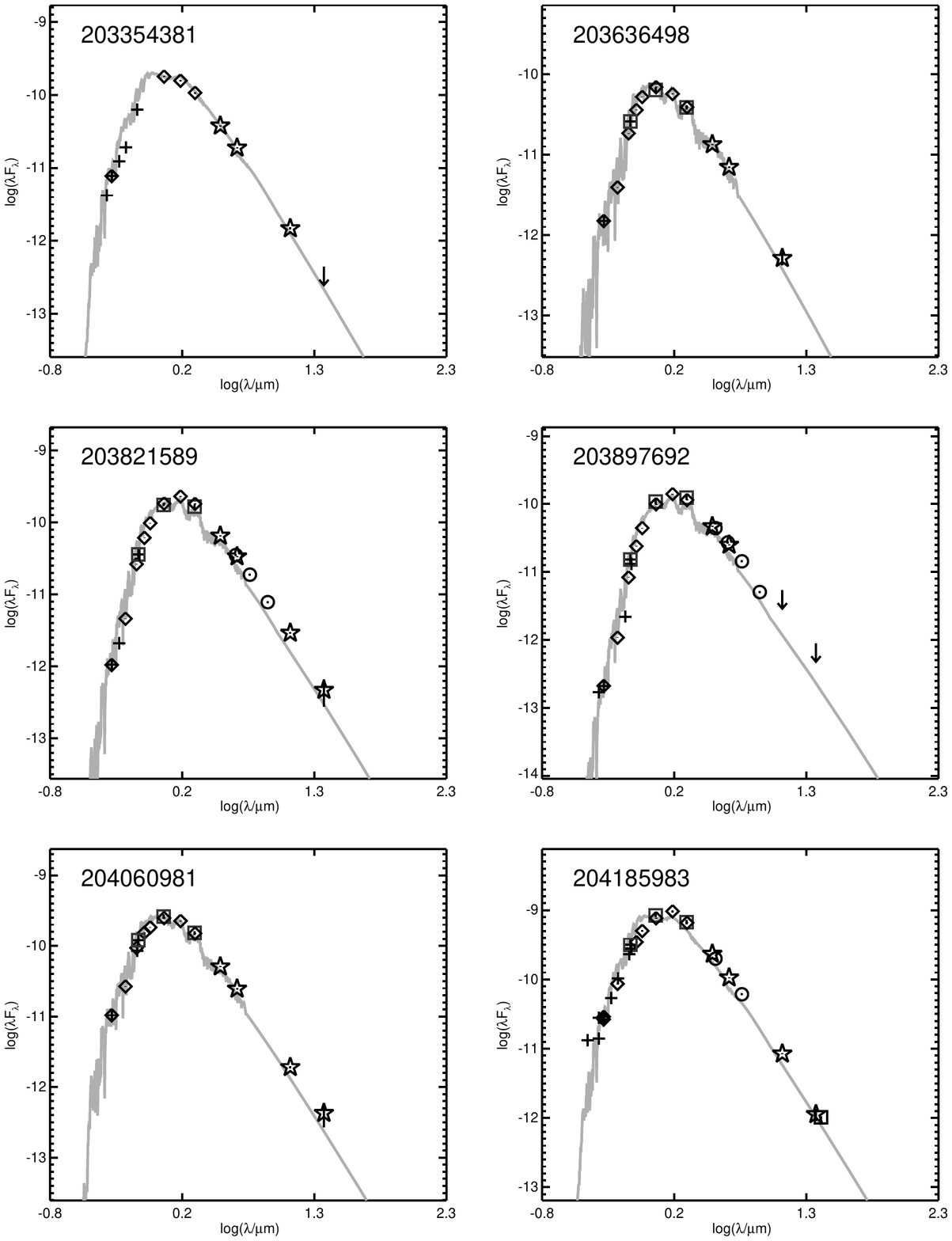}
\caption{Spectral energy distributions for six of the stars in Table 1.
See the text for a description of the symbols.
\label{fig:SEDS1}}
\end{figure}

\begin{figure}[ht]
\epsscale{0.65}
\plotone{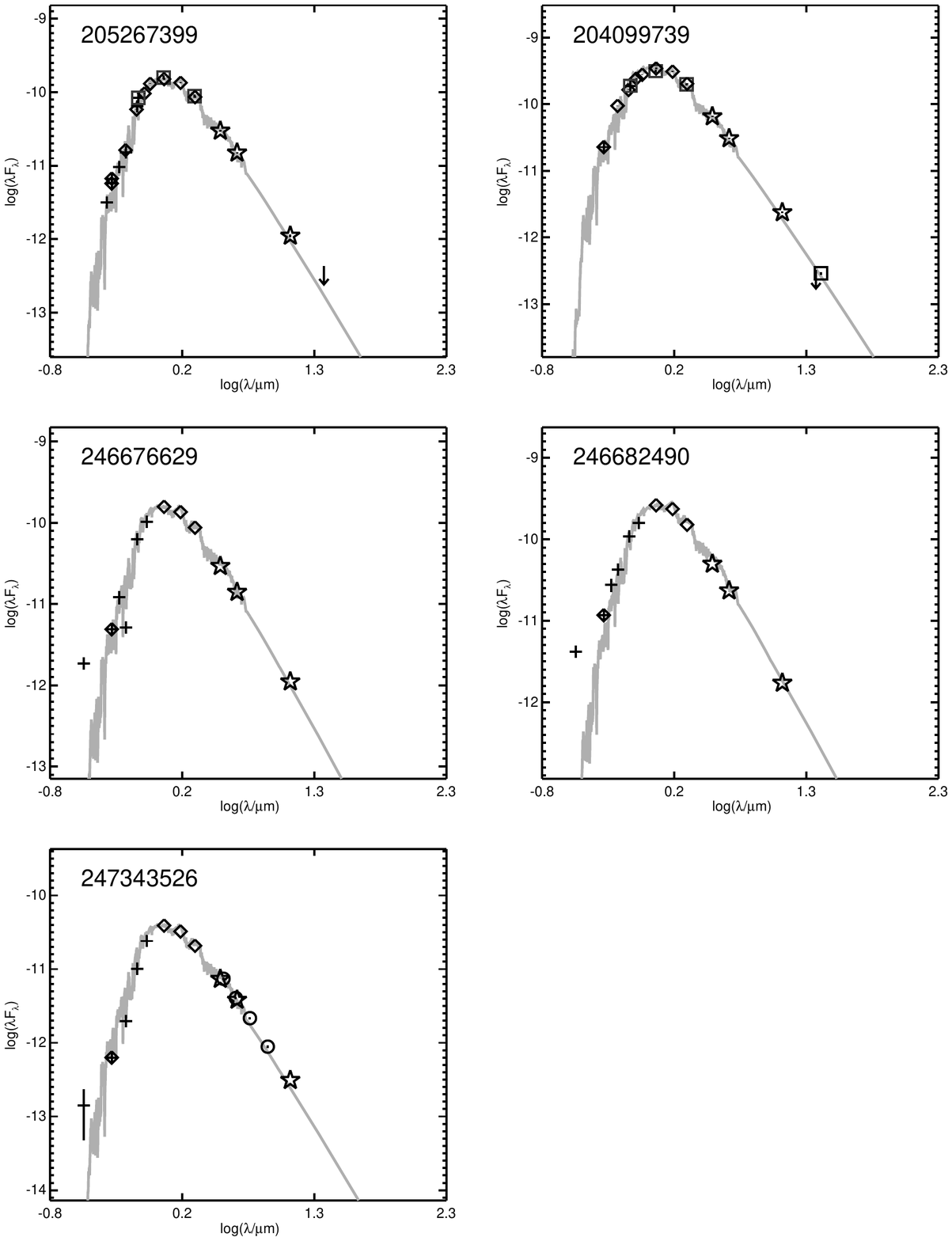}
\caption{Spectral energy distributions for the remaining stars in Table 1.
\label{fig:SEDS2}}
\end{figure}

Spectral energy distributions for all the stars from Table 1 are
provided here as Figure \ref{fig:SEDS1} and Figure \ref{fig:SEDS2}.
Plots are log $\lambda F_{\lambda}$ in cgs units (ergs s$^{-1}$
cm$^{-2}$) against log $\lambda$ in microns. Symbols: $+$: optical
literature (SDSS, APASS); box at short bands: DENIS $IJK$; diamond:
2MASS $JHK_s$; circle: IRAC; box at long bands: MIPS; stars: WISE;
arrows: limits; vertical bars (often smaller than the symbol) denote
uncertainties.  A Kurucz-Lejeune model for the corresponding spectral
type (taken to be M3 for the one star without known type) is also shown as
the grey line, normalized to the observations at $K_s$. Note that this
is not a robust fit, but just to ``guide the eye.'' The $\rho$\ Oph
members have
been lightly reddened to better fit the optical points ($A_J \leq 0.7$
in all cases).  Essentially all the stars have SEDs consistent with  pure
photospheres; none have large IR excesses.

\end{document}